\documentclass[10pt]{article}
\usepackage[cp1251]{inputenc}
\usepackage[dvips]{epsfig}   \usepackage[dvips]{changebar}
\usepackage[dvips]{graphicx}

\def\lsim{\
  \lower-1.2pt\vbox{\hbox{\rlap{$<$}\lower5pt\vbox{\hbox{$\sim$}}}}\ }
\def\gsim{\
  \lower-1.2pt\vbox{\hbox{\rlap{$>$}\lower5pt\vbox{\hbox{$\sim$}}}}\ }

\usepackage{amsmath,enumerate, amsfonts, amssymb,amsthm}

\begin{document}
\title{On a fragmented condensate in a uniform Bose system}
 \author{Maksim Tomchenko
\bigskip \\ {\small Bogolyubov Institute for Theoretical Physics} \\
 {\small 14b, Metrolohichna Str., Kyiv 03143, Ukraine} \\
  {\small E-mail:mtomchenko@bitp.kiev.ua}}
 \date{\empty}
 \maketitle
 \large
 \sloppy
\textit{ According to the well-known analysis by Nozi\'{e}res, the
fragmentation of the condensate increases the energy of a uniform
interacting Bose system. Therefore, at $T= 0$ the condensate should
be nonfragmented. We perform a more detailed analysis and show that
the result by Nozi\'{e}res is not general. We find that, in a dense
Bose system, the formation of a crystal-like structure with a
fragmented condensate is possible. The effect is related to a
nonzero size of real atoms. Moreover, the wave functions studied by
Nozi\'{e}res are not eigenfunctions of the Hamiltonian and,
therefore, do not allow one to judge with confidence about the
structure of the condensate in the ground state. We have constructed
the wave functions in such a way that they are eigenfunctions of the
Hamiltonian. The results show that the fragmentation of the
condensate (quasicondensate) is possible for a finite
one-dimensional uniform system at
low temperatures and a weak coupling. } \\
%Keywords: interacting bosons, fragmented condensate,
%quasicondensate.

\section{Introduction}
\label{intro} The Bose--Einstein condensation (BEC) is a beautiful
purely quantum property \cite{einstein1924,bog1947,penronz}. The
early history of the ideas on a condensate can be found in review
\cite{griffin1999}. BEC in gases and fluids is intensively studied
experimentally and theoretically
\cite{cornell1996,ketterle1996,leggett2006,pit2006,pethick2008,lopes2017}.
However, some open questions remain in this field. In particular, in
addition to the one-particle condensate, the two-particle condensate
can exist in a Bose system with repulsive interaction
\cite{coniglio1967,coniglio1969,ristig1976,ristig1978,nozieres1982,mt2006,mathey2009}.
It is not quite clear whether the existence of the three-particle
and higher condensates is possible. According to the calculation
with regard for the two- and three-particle correlations, the
three-particle and higher condensates are absent in a
three-dimensional (3D) Bose liquid \cite{mt2006}.

Of high interest is also the question whether a condensate can be
fragmented. The condensate in a stationary system of $N$ identical
structureless bosons is called fragmented \cite{leggett2006}, if the
diagonal expansion of the single-particle density matrix
\begin{equation}
F_{1}(\textbf{r},\textbf{r}^{\prime})  =
\sum\limits\limits_{j=1}^{\infty}\lambda_{j}\phi^{*}_{j}(\textbf{r}^{\prime})\phi_{j}(\textbf{r})
     \label{n2} \end{equation}
contains two or more macroscopic natural occupations $\lambda_{j}$:
for example,  $\lambda_{1}, \lambda_{2} \sim N$. Here, the natural
orbitals $\phi_{j}(\textbf{r})$ form the complete collection of
orthonormal functions, and $\lambda_{j}$ are the occupation numbers
of the single-particle states $\phi_{j}(\textbf{r})$. We use the
normalization of the function
$F_{1}(\textbf{r},\textbf{r}^{\prime})$, for which
$\lambda_{1}+\ldots +\lambda_{\infty}= N$. Pollock
\cite{pollock1967} and Nozi\'{e}res \cite{nozieres1995} argued that
the energy $E^{(2)}$ of a uniform system with two condensates should
be higher than the energy $E^{(1)}$ of a system with one condensate.
Indeed, for the repulsive point interaction
$U(|\textbf{r}_{j}-\textbf{r}_{l}|)=2c\delta(\textbf{r}_{j}-\textbf{r}_{l})$
the difference $E^{(2)}-E^{(1)}$ is close to the exchange energy
 \cite{nozieres1995,spekkens1998}:
\begin{equation}
E^{(2)}-E^{(1)}\simeq
2cN_{1}N_{2}\int\phi^{2}_{1}(\textbf{r})\phi^{2}_{2}(\textbf{r})d\textbf{r}>0.
     \label{0-2} \end{equation}
Here, we assume the following: All $N$ atoms of the system with one
condensate are in the state $\phi_{0}(\textbf{r})$. For the system
with two condensates, $N_{1}$ atoms are in the state
$\phi_{1}(\textbf{r})$, $N_{2}$ atoms occupy the state
$\phi_{2}(\textbf{r})$, $N_{1}+N_{2}=N$, and
$\phi^{2}_{1}(\textbf{r})\simeq \phi^{2}_{2}(\textbf{r})\simeq
\phi^{2}_{0}(\textbf{r})$. In this case, the fragmentation of the
condensate costs a macroscopic energy
\cite{pollock1967,nozieres1995}. If the condensates are separated in
the $\textbf{r}$-space, then the overlapping of the functions
$\phi_{1}(\textbf{r})$ and $\phi_{2}(\textbf{r})$ is small.
Therefore, to find the value of $E^{(2)}-E^{(1)},$ it is necessary
to consider additional terms. The analysis shows that, for the Bose
gas in a double-well potential of a trap, the state with two
condensates, which are localized at different minima of a trap, is
energy-gained \cite{spekkens1998,spekkens1999}. The other examples
of a fragmented condensate  can be found in
\cite{leggett2006,baym2006}. The solutions with a fragmented
condensate were obtained for one-dimensional (1D) and
two-dimensional (2D) Bose gases in a trap
\cite{sakmann2009,fischer2010,zollner2010,streltsov2013,sakmann2014,lode2014,weiner2017,fischer2017}.
The fragmentation of the condensate of quasiparticles is discussed
in review \cite{combescot2017}.

In the present work, we will analyze the problem of the
fragmentation of the condensate in more details than in
\cite{pollock1967,nozieres1995}. We will show that the fragmentation
of the condensate is possible even for a \textit{uniform} system
(analogous result was obtained previously \cite{mtjltp2016} without
general analysis of the problem of fragmentation). In this case, the
condensates are not separated in the $\textbf{r}$-space, in contrast
to the solutions in
\cite{spekkens1998,spekkens1999,sakmann2009,fischer2010,zollner2010,streltsov2013,sakmann2014,lode2014,fischer2017}.
We will consider the problem step by step, by passing from a more
crude description to an accurate one. In Sections 2 and 3, we will
show that the approach by Pollock-Nozi\'{e}res
\cite{pollock1967,nozieres1995} has two weak places: point
interatomic potential and Hartree--Fock wave functions. We will see
that the use of a nonpoint potential leads to the possibility of a
crystal-like solution with fragmented condensate (Sect. 2). The
transition from Hartree--Fock wave functions to the more accurate
collective description is considered in Sect. 3. The solutions with
fragmented condensate in Sections 2 and 3 are approximate. In Sect.
4,
%with the help of the collective approach
we will find the accurate solution for a fragmented condensate in
the 1D Bose gas.

\section{Periodic Bose system: quasi-single-particle  approach}
\label{sec:2} In this section, we will carry on the analysis similar
to the analysis by Pollock \cite{pollock1967} and by Nozi\'{e}res
\cite{nozieres1995} and will take into account the nonpointness
(nonzero interaction radius) of real particles. Consider the
periodic system of  $N$ bosons with repulsive interaction
($\nu(0)>0$). The exact Hamiltonian of the system reads
\begin{eqnarray}
 \hat{H}&=& -\frac{\hbar^2}{2m}\int\limits_{V}d\textbf{r}\hat{\psi}^{+}(\textbf{r},t)\triangle \hat{\psi}(\textbf{r},t)
 \nonumber \\ &+&
 \frac{1}{2}\int\limits_{V}d\textbf{r} d\textbf{r}^{\prime}U(|\textbf{r}-\textbf{r}^{\prime}|)
 \hat{\psi}^{+}(\textbf{r},t)\hat{\psi}^{+}(\textbf{r}^{\prime},t)\hat{\psi}(\textbf{r},t)\hat{\psi}(\textbf{r}^{\prime},t),
     \label{f-0} \end{eqnarray}
\begin{equation}
 U(|\textbf{r}-\textbf{r}^{\prime}|) = \frac{1}{V}
 \sum\limits_{\textbf{k}}\nu(\textbf{k})e^{i\textbf{k}(\textbf{r}-\textbf{r}^{\prime})},
     \label{p} \end{equation}
where  $\textbf{k}=2\pi \left (\frac{j_{x}}{L_{x}},
\frac{j_{y}}{L_{y}}, \frac{j_{z}}{L_{z}} \right )$,
 $j_{x}, j_{y}, j_{z}=0, \pm 1, \pm 2, \ldots$,  $L_{x}, L_{y}, L_{z}$ are
the sizes of the system, and $V=L_{x}L_{y}L_{z}$. In this section,
we consider an isolated quantum-mechanical system, being in some
pure state $\Psi(\textbf{r}_{1},\ldots \textbf{r}_{N})$. In view of
this, we use the quantum-mechanical average \cite{land3}: $\langle
\hat{A} \rangle=\int d\textbf{r}_{1}\ldots
d\textbf{r}_{N}\Psi^{*}\hat{A}\Psi$.

\subsection{Solutions with one, two, and three condensates}
If all atoms are in \textit{one} condensate of atoms with zero
momentum, then we have the wave function of the system
\begin{eqnarray}
\Psi=C_{1}(\hat{a}^{+}_{0})^{N}|vac\rangle,
     \label{f-n1} \end{eqnarray}
the second-quantized operator
\begin{eqnarray}
\hat{\psi}(\textbf{r},t)  = \hat{a}_{0}/\sqrt{V},
     \label{f-n2} \end{eqnarray}
and $\hat{a}^{+}_{0}\hat{a}_{0}=\hat{N}$. In this case,
\begin{eqnarray}
\hat{H}^{(1)}=\frac{\nu(0)(\hat{N}^{2}-\hat{N})}{2V}, \quad E^{(1)}=
\langle \hat{H}^{(1)} \rangle= \frac{\nu(0)(N^{2}-N)}{2V},
     \label{f-1} \end{eqnarray}
where $E^{(1)}$ is the energy of the system. Let the atoms be
distributed over \textit{three} states:
\begin{eqnarray}
\Psi=C_{3}(\hat{a}^{+}_{0})^{N_{0}}\cdot(\hat{a}^{+}_{\textbf{k}})^{N_{\textbf{k}}}\cdot(\hat{a}^{+}_{-\textbf{k}})^{N_{-\textbf{k}}}|vac\rangle,
     \label{f-n3} \end{eqnarray}
\begin{eqnarray}
\hat{\psi}(\textbf{r},t)
=V^{-1/2}(\hat{a}_{0}+\hat{a}_{\textbf{k}}e^{i\textbf{k}\textbf{r}}+\hat{a}_{-\textbf{k}}e^{-i\textbf{k}\textbf{r}}),
     \label{f-n4} \end{eqnarray}
$\hat{a}^{+}_{0}\hat{a}_{0}=\hat{N}_{0}$,
$\hat{a}^{+}_{\textbf{k}}\hat{a}_{\textbf{k}}=\hat{N}_{\textbf{k}}$,
$\hat{a}^{+}_{-\textbf{k}}\hat{a}_{-\textbf{k}}=\hat{N}_{-\textbf{k}}$,
$\hat{N}_{0}+\hat{N}_{\textbf{k}}+\hat{N}_{-\textbf{k}}=\hat{N}$ (it
is seen from the analysis by Bogoliubov \cite{bog1947} that the
states $e^{i\textbf{k}\textbf{r}}$ and $e^{-i\textbf{k}\textbf{r}}$
are coupled [this is indicated by terms
$\hat{b}^{+}_{\textbf{k}}\hat{b}^{+}_{-\textbf{k}}$ and
$\hat{b}_{\textbf{k}}\hat{b}_{-\textbf{k}}$ in Eq. (\ref{f-4})
below]; therefore, we consider them together).  In this case,
$N_{0}=\langle \hat{N}_{0} \rangle$, $N_{\textbf{k}}=\langle
\hat{N}_{\textbf{k}} \rangle$, $N_{-\textbf{k}}=\langle
\hat{N}_{-\textbf{k}} \rangle$. The numbers $N_{\textbf{k}}$ and
$N_{-\textbf{k}}$ can be macroscopic or microscopic. Then
\begin{eqnarray}
F_{1}(\textbf{r},\textbf{r}^{\prime})=\langle
\hat{\psi}^{+}(\textbf{r}^{\prime},t)\hat{\psi}(\textbf{r},t)
\rangle
=N_{0}\frac{1}{V}+N_{\textbf{k}}\frac{e^{i\textbf{k}(\textbf{r}-\textbf{r}^{\prime})}}{V}+
N_{-\textbf{k}}\frac{e^{-i\textbf{k}(\textbf{r}-\textbf{r}^{\prime})}}{V}.
     \label{f-n5} \end{eqnarray}
We have obtained the diagonal expansion (\ref{n2}) with
$\lambda_{0}=N_{0}$, $\lambda_{\textbf{k}}=N_{\textbf{k}}$,  and
$\lambda_{-\textbf{k}}=N_{-\textbf{k}}$. That is, the definition of
a fragmented condensate on the basis of formulae like (\ref{f-n3}),
(\ref{f-n4}) is equivalent to that on the basis of (\ref{n2}).

In order to find the Hamiltonian (\ref{f-0}) with the operator
$\hat{\psi}(\textbf{r},t)$ (\ref{f-n4}), we should take into account
in the operator
$\hat{\psi}^{+}(\textbf{r},t)\hat{\psi}^{+}(\textbf{r}^{\prime},t)
\hat{\psi}(\textbf{r},t)\hat{\psi}(\textbf{r}^{\prime},t)$ the terms
\begin{eqnarray}
&&\frac{1}{V^{2}}\left \{
\hat{a}^{+}_{\textbf{k}}\hat{a}^{+}_{\textbf{k}}\hat{a}_{\textbf{k}}\hat{a}_{\textbf{k}}+
\hat{a}^{+}_{-\textbf{k}}\hat{a}^{+}_{-\textbf{k}}\hat{a}_{-\textbf{k}}\hat{a}_{-\textbf{k}}\right.\nonumber
\\&&+
\hat{a}^{+}_{\textbf{k}}\hat{a}^{+}_{-\textbf{k}}\hat{a}_{\textbf{k}}\hat{a}_{-\textbf{k}}\left
(e^{i2\textbf{k}(\textbf{r}-\textbf{r}^{\prime})}+
e^{-i2\textbf{k}(\textbf{r}-\textbf{r}^{\prime})}+2\right )
\nonumber
\\ &&+ \hat{a}^{+}_{0}\hat{a}^{+}_{0}\hat{a}_{0}\hat{a}_{0}+ \hat{a}^{+}_{0}\hat{a}_{0}
(\hat{a}^{+}_{\textbf{k}}\hat{a}_{\textbf{k}}+\hat{a}^{+}_{-\textbf{k}}\hat{a}_{-\textbf{k}})\left
(e^{i\textbf{k}(\textbf{r}-\textbf{r}^{\prime})}+
e^{-i\textbf{k}(\textbf{r}-\textbf{r}^{\prime})}+2\right ) \nonumber
\\ &&+ \left. (\hat{a}^{+}_{0}\hat{a}^{+}_{0}
\hat{a}_{\textbf{k}}\hat{a}_{-\textbf{k}}+\hat{a}^{+}_{\textbf{k}}\hat{a}^{+}_{-\textbf{k}}\hat{a}_{0}\hat{a}_{0})\left
(e^{i\textbf{k}(\textbf{r}-\textbf{r}^{\prime})}+
e^{-i\textbf{k}(\textbf{r}-\textbf{r}^{\prime})}\right ) \right \}.
     \label{f-2} \end{eqnarray}
Using the formulae
$\hat{a}^{+}_{0}\hat{a}^{+}_{0}\hat{a}_{0}\hat{a}_{0}=\hat{N}^{2}_{0}-\hat{N}_{0}$,
$\nu(-\textbf{k})=\nu(\textbf{k})=\nu(k)$ and Eqs. (\ref{f-0}),
(\ref{p}),  (\ref{f-2}), we get the Hamiltonian and the energy of
the system:
\begin{equation}
\hat{H}^{(1+2)}=\hat{H}^{(1)}+(\hat{N}_{\textbf{k}}+\hat{N}_{-\textbf{k}})\left
[K(k)+\frac{\hat{N}_{0}\nu(k)}{V}\right ]+
\frac{\nu(2k)}{V}\hat{N}_{\textbf{k}}\hat{N}_{-\textbf{k}}+\hat{H}_{scat},
     \label{f-3} \end{equation}
\begin{equation}
\hat{H}_{scat}= \frac{\nu(k)}{V}(\hat{a}^{+}_{0}\hat{a}^{+}_{0}
\hat{a}_{\textbf{k}}\hat{a}_{-\textbf{k}}+\hat{a}^{+}_{\textbf{k}}\hat{a}^{+}_{-\textbf{k}}\hat{a}_{0}\hat{a}_{0}),
     \label{f-3b} \end{equation}
\begin{equation}
E^{(1+2)}=\langle \hat{H}^{(1+2)} \rangle=
E^{(1)}+(N_{\textbf{k}}+N_{-\textbf{k}})[K(k)+n_{0}\nu(k)]+N_{\textbf{k}}N_{-\textbf{k}}\nu(2k)/V,
     \label{f-3e} \end{equation}
where $n_{0}=\frac{N_{0}}{V}$,
$K(k)=\frac{\hbar^{2}\textbf{k}^{2}}{2m}$. This three-condensate
solution $(1+2)$ yields immediately two \textit{two-condensate}
solutions. We set $N_{0}=0$ and $N_{-\textbf{k}}=N-N_{\textbf{k}}$.
Then the solution $(1+2)$ transits to the solution $(0+2)$:
\begin{eqnarray}
\Psi=C_{02}(\hat{a}^{+}_{\textbf{k}})^{N_{\textbf{k}}}\cdot(\hat{a}^{+}_{-\textbf{k}})^{N_{-\textbf{k}}}|vac\rangle,
     \label{f-n6} \end{eqnarray}
\begin{eqnarray}
\hat{\psi}(\textbf{r},t)
=V^{-1/2}(\hat{a}_{\textbf{k}}e^{i\textbf{k}\textbf{r}}+\hat{a}_{-\textbf{k}}e^{-i\textbf{k}\textbf{r}}),
     \label{f-n7} \end{eqnarray}
\begin{equation}
\hat{H}^{(0+2)}=\hat{H}^{(1)}+(\hat{N}_{\textbf{k}}+\hat{N}_{-\textbf{k}})K(k)+
\frac{\nu(2k)}{V}\hat{N}_{\textbf{k}}\hat{N}_{-\textbf{k}},
     \label{f-n8} \end{equation}
\begin{equation}
E^{(0+2)}= E^{(1)}+NK(k)+N_{\textbf{k}}(N-N_{\textbf{k}})\nu(2k)/V.
\label{f-n9} \end{equation}

If we set $N_{-\textbf{k}}=0$ and $N_{0}=N-N_{\textbf{k}}$ in the
three-condensate solution, we find another solution with two
condensates:
\begin{eqnarray}
\Psi=C_{11}(\hat{a}^{+}_{0})^{N_{0}}\cdot(\hat{a}^{+}_{\textbf{k}})^{N_{\textbf{k}}}|vac\rangle,
     \label{f-n10} \end{eqnarray}
\begin{eqnarray}
\hat{\psi}(\textbf{r},t)
=V^{-1/2}(\hat{a}_{0}+\hat{a}_{\textbf{k}}e^{i\textbf{k}\textbf{r}}),
     \label{f-n11} \end{eqnarray}
\begin{equation}
\hat{H}^{(1+1)}=\hat{H}^{(1)}+\hat{N}_{\textbf{k}}\left
[K(k)+\frac{\hat{N}_{0}\nu(k)}{V}\right ],
     \label{f-n12} \end{equation}
\begin{equation}
E^{(1+1)}=
E^{(1)}+N_{\textbf{k}}[K(k)+n\nu(k)]-N^{2}_{\textbf{k}}\nu(k)/V.
     \label{f-n13} \end{equation}
We note that formulae (\ref{f-n10})--(\ref{f-n13}), written in a
different form, were previously obtained by Pollock
\cite{pollock1967}. Work \cite{pollock1967} is little known, but it
contains the Nozi\'{e}res' result and was published much earlier
than the work by Nozi\'{e}res \cite{nozieres1995}. Formulae
(\ref{f-3})--(\ref{f-n13})  allow us to make some interesting
conclusions.

\subsection{Analysis of solutions: when is the fragmentation possible?}
For $K(k)\approx 0$ and $\nu(2k)=\nu(k)=\nu(0)>0,$ we obtain
$E^{(1+2)}\approx
E^{(1)}+(N_{\textbf{k}}+N_{-\textbf{k}})n_{0}\nu(0)+N_{\textbf{k}}N_{-\textbf{k}}\nu(0)/V>E^{(1)}$.
Thus, we arrive at the Pollock-Nozi\'{e}res' conclusion
\cite{pollock1967,nozieres1995}: the fragmentation of the condensate
increases the energy of the system. If $N_{0}=0$ or
$N_{-\textbf{k}}=0$, the conclusion is the same. However, the
equality $\nu(k)=\nu(0)$ holds at any $k$ only for the point
interaction. As known, the point potential allows one to properly
describe the long-wave properties of a system. Below, we will get
solutions with fragmented condensate, for which the fragments
$N_{\textbf{k}}, N_{-\textbf{k}}$ of a condensate are short-wave
solutions. In order to properly describe the short-wave properties
of a system, we need to use a \textit{nonpoint} potential. Indeed,
any real interatomic potential has a nonzero radius $r_{0}\sim
1\,\mbox{\AA}$. In this case,
 $\nu(k)\sim -0.1\nu(0)<0$ at $k\sim  \pi/r_{0}$. The real
potentials have a complicated form (for $^4$He-atoms, see
\cite{aziz1991,szalewicz2012}). Very approximately, we can consider
an atom as a semitransparent ball:
 \begin{equation}
 U(\textbf{r}) \approx
\left [ \begin{array}{ccc}
    U_{0}>0  & \   r\leq d_{0},  & \\
    0  & \ r>d_{0}, &
\label{p-1} \end{array} \right. \end{equation} where
$d_{0}=2r_{0}\approx 2$--$3\,\mbox{\AA}$,  $U_{0}\sim
10^{3}$--$10^{6}\,$K. We note that the simple model potential
(\ref{p-1}) allows us to qualitatively correctly reproduce the
behavior of the Fourier-transform $\nu(k)$ of a real complicated
potential. In the 3D case, the Fourier transform of the potential
(\ref{p-1}) is
\begin{eqnarray}
 \nu(k) = \int\limits_{-L_{x}}^{L_{x}}dx \int\limits_{-L_{y}}^{L_{y}}dy
 \int\limits_{-L_{z}}^{L_{z}}dz U(\textbf{r})e^{-i\textbf{k}\textbf{r}} &=& 4\pi U_{0}d_{0}^{3}f_{3}(kd_{0}),
      \label{p-2} \end{eqnarray}
where $f_{3}(g)=(\sin{g}-g\cos{g})/g^{3}$.  In the 1D case, we have
\begin{eqnarray}
 \nu(k)  &=& 2 U_{0}d_{0}f_{1}(kd_{0}), \quad
 f_{1}(g)=\frac{\sin{g}}{g}.
      \label{p-1D} \end{eqnarray}
The functions $f_{1}(g)$ and $f_{3}(g)$ are oscillatory (see Fig.
1).

\begin{figure*}
\includegraphics[width=.6\textwidth]{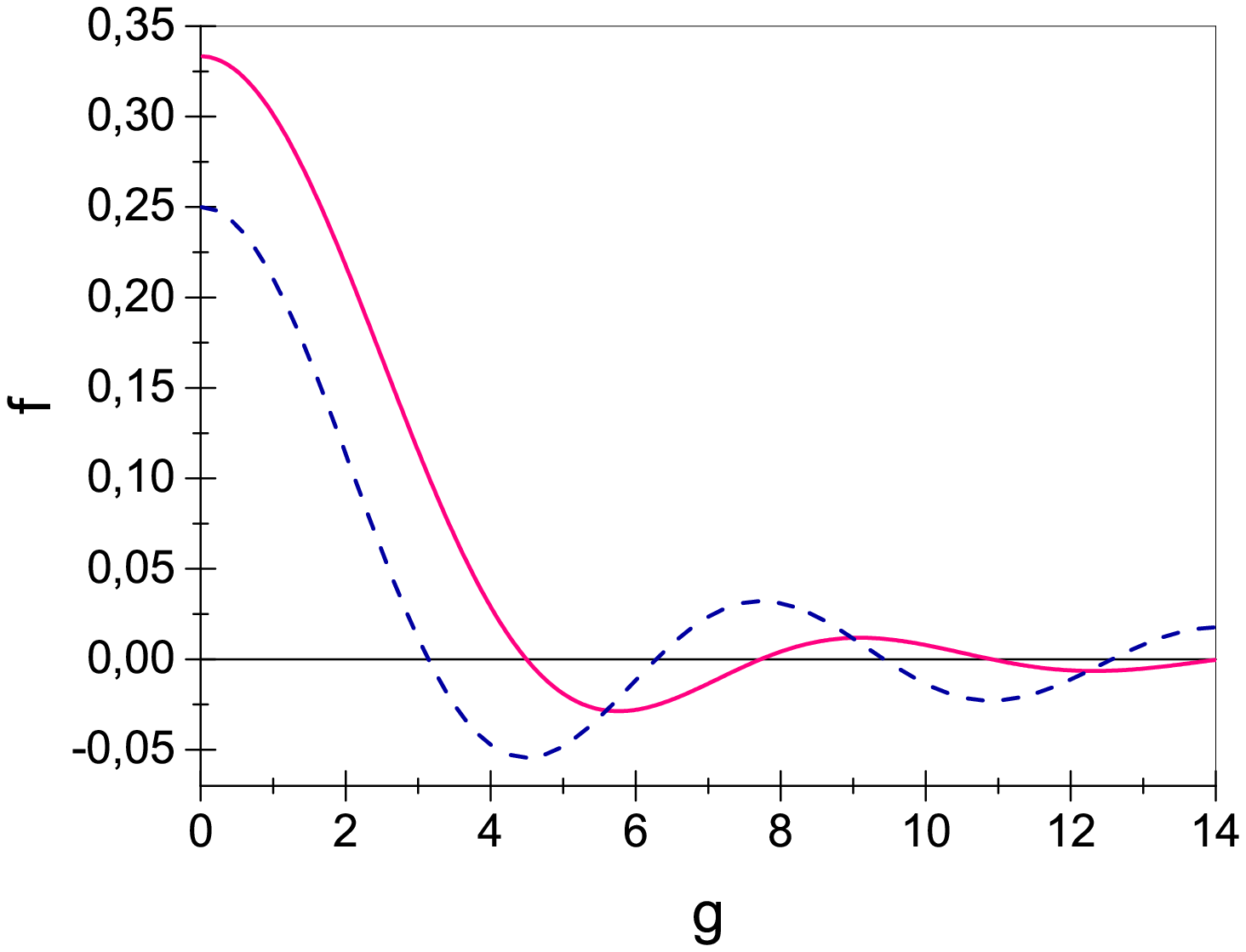}
\caption{[Color online] The functions $f_{1}(g)$ (dashed line) and
$f_{3}(g)$ (solid line). The function $f_{1}(g)$ is multiplied by
$1/4$.}
          \label{fig:1}                  \end{figure*}

If the values of $k$ lie near
% $k_{0}$  (the coordinate of
the first minimum of the function $\nu(k)$ and if $n_{0}$ is large,
we have $K(k)+n_{0}\nu(k)<0$. Then it is seen from Eq. (\ref{f-3e})
that the relation $E^{(1+2)}<E^{(1)}$ becomes possible. In this
case, the average value of the energy of the state $(1+2)$ is less
than for the state $1$ (with one condensate). Therefore, the
\textit{fragmentation of the condensate is possible}. If
$N_{-\textbf{k}}=0$ or $N_{0}=0$, the conclusion is the same. Note
that the considered states are uniform. In particular, for the state
with three condensates, the particle number density $n(\textbf{r})$
is constant:
\begin{eqnarray}
n(\textbf{r})=\langle
\hat{\psi}^{+}(\textbf{r},t)\hat{\psi}(\textbf{r},t)
\rangle=V^{-1}\langle \hat{a}^{+}_{0}\hat{a}_{0}+
\hat{a}^{+}_{\textbf{k}}\hat{a}_{\textbf{k}}+\hat{a}^{+}_{-\textbf{k}}\hat{a}_{-\textbf{k}}
\rangle=N/V.
     \label{ro} \end{eqnarray}

Consider the conditions, under which the fragmentation is possible,
in more details. In order to determine the smallest value of the
function $E^{(1+2)}(N_{\textbf{k}}, N_{-\textbf{k}})$ (\ref{f-3e}),
we need to find the minimum of this function in the internal domain
of the phase space ($0<N_{\textbf{k}}, N_{-\textbf{k}}<N$;
$N_{\textbf{k}}+ N_{-\textbf{k}}<N$) and the boundary values of the
function (one boundary corresponds to $N_{-\textbf{k}}=0$, and
another one is set by the equality $N_{\textbf{k}}+
N_{-\textbf{k}}=N$). The extremum corresponds to
\begin{equation}
N_{\textbf{k}}=N_{-\textbf{k}}=\frac{K(k)+n\nu(k)}{4n\nu(k)-n\nu(2k)}N.
\label{nmin1} \end{equation} In this case,
\begin{eqnarray}
E^{(1+2)}=E^{(1)}+N_{\textbf{k}}(K(k)+n\nu(k)).
     \label{f-n14} \end{eqnarray}
It is a minimum, if $\nu(k)<0$. We see that $E^{(1+2)}< E^{(1)}$, if
 $K(k)+n\nu(k)<0$. Next, we consider the boundary region
$N_{\textbf{k}}+ N_{-\textbf{k}}=N$, which is equivalent to the
analysis of the above-presented solution $(0+2)$. We need to
determine a minimum of the function $E^{(0+2)}(N_{\textbf{k}})$ at
$0< N_{\textbf{k}}<N$ and to compare it with the boundary value
$E^{(0+2)}(N_{\textbf{k}}=0)=E^{(1)}+NK(k)$. The minimum corresponds
to the relations $N_{\textbf{k}}=N_{-\textbf{k}}=N/2$, $\nu(2k)<0$.
At this point of the minimum,
\begin{eqnarray}
E^{(0+2)}= E^{(1)}+N(K(2k)+n\nu(2k))/4.
     \label{f-n15} \end{eqnarray}
This value is less than the energies $E^{(1)}+NK(k)$ and $E^{(1)}$,
if $K(2k)+n\nu(2k)<0$. Eventually, we study another boundary region
of function (\ref{f-3e}): $N_{-\textbf{k}}=0$. This is equivalent to
the analysis of the solution $(1+1)$ obtained in  \cite{pollock1967}
and above. The energy $E^{(1+1)}(N_{\textbf{k}})$ has a minimum at
\begin{equation}
\frac{N_{\textbf{k}}}{N}=\frac{K(k)+n\nu(k)}{2n\nu(k)} \label{nmin2}
\end{equation}
and $\nu(k)<0$. This implies that the solution with
$N_{\textbf{k}}>0$ exists at $K(k)+n\nu(k)<0$. At the minimum, we
have
\begin{eqnarray}
E^{(1+1)}(N_{\textbf{k}})=E^{(1)}+N\frac{(K(k)+n\nu(k))^{2}}{4n\nu(k)}.
     \label{f-n16} \end{eqnarray}
If $\nu(k)<0$, we obtain $E^{(1+1)}< E^{(1)}$. On the edges
($N_{\textbf{k}}=0; N$) the energy is higher: $E^{(1+1)}= E^{(1)};
E^{(1)}+NK(k)$.

\begin{figure*}
\includegraphics[width=.6\textwidth]{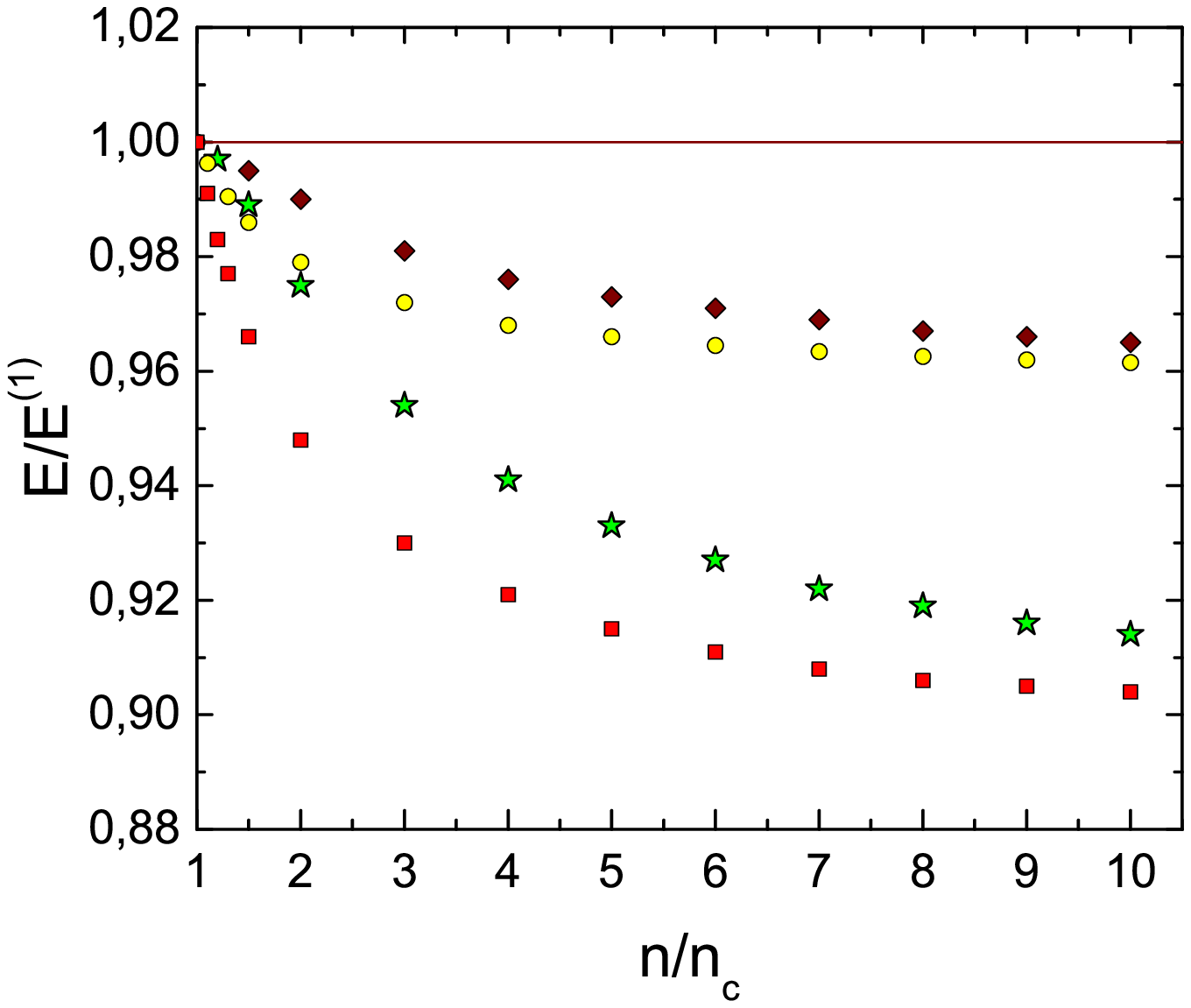}
\caption{[Color online] Smallest values of the function
$E^{(1+2)}(N_{\textbf{k}}, N_{-\textbf{k}}, k)/E^{(1)}$ at the given
density $n$ in the $1D$  (squares) and  $3D$ (circles) cases. They
are found numerically from Eqs. (\ref{f-1}), (\ref{f-3e}) and
$N_{0}=N-N_{\textbf{k}}-N_{-\textbf{k}}$ for the potentials
(\ref{p-2}), (\ref{p-1D}) and all possible values of
$N_{\textbf{k}}$, $N_{-\textbf{k}},$ and $k$ ($0<k<\infty$, $0\leq
N_{\textbf{k}}\leq N$, $0\leq N_{-\textbf{k}}\leq N$ under the
condition $N_{\textbf{k}}+N_{-\textbf{k}}\leq N$). The values of
$n_{c}$ in the $1D$ and $3D $ cases are presented in the text. We
also show smallest values of $E^{(1+1)}(N_{\textbf{k}}, k)/E^{(1)}$
for the given $n$ in the $1D$ (stars) and  $3D$ (rhombs) cases,
determined from Eq. (\ref{f-n16}). At $n/n_{c}\leq 1$ the smallest
$E^{(1+2)}(N_{\textbf{k}}, N_{-\textbf{k}}, k)$ and
$E^{(1+1)}(N_{\textbf{k}}, k)$ are equal to $E^{(1)}$.}
          \label{fig:2}                  \end{figure*}

Thus, in all three cases, we obtain the condition $n>n_{c}$, where
the critical density $n_{c}$ is the smallest positive density, for
which the equality $K(k)+n\nu(k)=0$ holds at some $k$. We found
numerically that $n_{c}\approx 84.2C_{1D}/d_{0}$, $g_{c}\approx
4.0781$ in the $1D$ case, and $n_{c}\approx
1091.45C_{3D}/d_{0}^{3}$, $g_{c}\approx 5.4486$ in the $3D$ case.
Here, $C_{1D}=\frac{\hbar^{2}}{4mU_{0}d_{0}^{2}}$,
$C_{3D}=\frac{C_{1D}}{2\pi}$, and $g_{c}$ is the value of
$g=kd_{0}$, for which the equality $K(k)+n\nu(k)=0$ yields
$n=n_{c}$.

We obtained numerically the smallest value of the energy
$E^{(1+2)}(N_{\textbf{k}}, N_{-\textbf{k}}, k)$ (\ref{f-3e}) as a
function of $N_{\textbf{k}}$, $N_{-\textbf{k}}$, $k$ at a fixed
$N=N_{0}+N_{\textbf{k}}+N_{-\textbf{k}}$ in the $3D$ and $1D$ cases,
by using the potentials (\ref{p-2}) and (\ref{p-1D}), respectively.
The analysis shows that at $n\leq n_{c}$ the smallest
$E^{(1+2)}(N_{\textbf{k}}, N_{-\textbf{k}}, k)$ corresponds to
$N_{\textbf{k}}=N_{-\textbf{k}}=0$. In this case, $N_{0}=N$,
$E^{(1+2)}(N_{\textbf{k}}, N_{-\textbf{k}}, k)$ coincides with
$E^{(1)},$ and the fragmentation is absent. At $n>n_{c}$ the
smallest $E^{(1+2)}(N_{\textbf{k}}, N_{-\textbf{k}}, k)$ is less
than $E^{(1)}$ and \textit{coincides with the energy $E^{(0+2)}$}
with $k\approx k_{c}=\frac{g_{c}}{2d_{0}}$. This  value of
$E^{(1+2)}$ is shown in Fig. 2.

Thus, at $n>n_{c}$ it is energy-gained for the state (\ref{f-n1}),
(\ref{f-n2}) with a single condensate to transit into the state
$(0+2)$ (\ref{f-n6}), (\ref{f-n7}) with two condensates
($N_{\textbf{k}}=N_{-\textbf{k}}=N/2$, $N_{0}=0$, condensate value
of $k$ depends weakly on $n$ and is close to $k_{c}/2$).

Note the following important point. In the  above solutions we
considered only a few $\textbf{k}$-harmonics in the operator
$\hat{\psi}$ and in the Hamiltonian. Of course, for the accurate
description of the system all $\textbf{k}$-harmonics should be taken
into account. Are the above obtained solutions $E^{(1+2)}$
(\ref{f-3e}), $E^{(0+2)}$ (\ref{f-n9}), and $E^{(1+1)}$
(\ref{f-n13}) close to the exact ones involving all
$\textbf{k}$-harmonics? We saw above that, at $n<n_{c},$ the state
(\ref{f-n1}) with one condensate is energy-gained. As an accurate
generalization of solution (\ref{f-n1})--(\ref{f-1}), we indicate
Bogoliubov's solution \cite{bog1947}. Under a weak coupling,
Bogoliubov ground-state energy $E_{0}$ is very close to $E^{(1)}$
(\ref{f-1}). In this case, function (\ref{f-n1}) is an eigenfunction
of the corresponding ``truncated'' Hamiltonian (\ref{f-1}).
Therefore, we suppose that if the wave function of the system
describes properly the structure of the condensate and is an
eigenfunction of the corresponding truncated Hamiltonian, and if the
coupling is weak or intermediate, then the corresponding
``truncated'' energy of the system is close to the exact
eigenenergy. In particular, the functions (\ref{f-n6}) and
(\ref{f-n10}) are eigenfunctions of the truncated Hamiltonians
(\ref{f-n8}) and (\ref{f-n12}), respectively. No accurate
generalization of solutions (\ref{f-n6}) and (\ref{f-n10}) has been
found. We expect that, for a weak and intermediate couplings,
energies (\ref{f-n9}) and (\ref{f-n13}) are close to the exact ones,
which can be determined in an accurate approach like Bogoliubov one
\cite{bog1947}. Note that the function (\ref{f-n3}) is not an
eigenfunction of the Hamiltonian $\hat{H}^{(1+2)}$ (\ref{f-3}) due
to the term $\hat{H}_{scat}$.

\subsection{Physical properties of solutions}
For real systems, the average distance $\bar{r}$ between atoms
should be larger than the atomic size: $\bar{r}\geq d_{0}$. The
strong overlapping of atoms ($\bar{r}\ll d_{0}$) is possible only at
very high external pressures; this case is omitted here.

We now make estimates for the $1D$ case. Let us introduce the
dimensionless Lieb-Liniger's parameter  \cite{ll1963}
$\gamma=\frac{m\nu(0)}{\hbar^{2}n}=\frac{1}{qC_{1D}^{2}2\cdot
84.2}$, where $q=n/n_{c}$. For $^{4}$He atoms, we have $d_{0}\simeq
2\,\mbox{\AA}$, then $C_{1D}\approx \frac{K\cdot k_{B}}{1.34 \cdot
U_{0}}$. The condition $\bar{r}=1/n\geq d_{0}$ yields the
inequalities $n=qn_{c}\leq 1/d_{0},$ $C_{1D}\leq 1/(q84.2)$, and
$\gamma \geq 42.1 q$. Since $q\geq 1$, we get $\gamma \gg 1$
corresponding to the strong coupling regime. For such $\gamma,$ the
solution for the ground-state energy is close to the solution for
impenetrable bosons ($\gamma = \infty$)
$E^{\infty}_{0}=\frac{N}{6}\frac{(\pi \hbar n)^{2}}{m} $
\cite{girardeau1960}. The relations
$\frac{E^{(0+2)}}{E^{\infty}_{0}}\simeq
\frac{E^{(1+1)}}{E^{\infty}_{0}}\simeq
\frac{E^{(1)}}{E^{\infty}_{0}}=\frac{1.5\bar{r}}{\pi^{2}d_{0}C_{1D}}\geq
\frac{1.5\cdot 84.2 q\bar{r}}{\pi^{2}d_{0}}\gsim 13$ imply that the
energies $E^{(0+2)}$ (\ref{f-n15}) and $E^{(1+1)}$ (\ref{f-n16}) are
much larger than the ground-state energy of a system of point bosons
with the same  $\nu(0)$. In other words, the above-considered states
$(0+2)$ and $(1+1)$ with two condensates are highly excited states
of the system.
%Such solutions are not of high interest, since
However, we are mainly interested in the structure of a condensate
for the \textit{ground} state.

In the $3D$ case, there are no exact solutions like
\cite{ll1963,girardeau1960}. Therefore, the estimates give less
information. From the above-presented formulae
$C_{3D}=\frac{C_{1D}}{2\pi}\approx \frac{K\cdot k_{B}}{2\pi\cdot
1.34 \cdot U_{0}}$ and $n_{c}\approx 1091.45C_{3D}/d_{0}^{3},$ we
get the critical density and the critical average interatomic
distance: $n_{c}\approx \frac{1}{d_{0}^{3}} \frac{130 K\cdot k_{B}}{
U_{0}}$, $\bar{r}_{c}=n^{-1/3}_{c}\approx \frac{d_{0}}{5}\left
(\frac{U_{0}}{K\cdot k_{B}}\right )^{1/3}$. The value of $U_{0}$ is
usually determined by means of fitting of a potential $U(r)$ to get
the best description of several experimental properties of a
substance. In addition, $U_{0}$ can be determined by means of the
calculation of the potential by the known structural factor $S(k)$.
For $^{4}$He atoms these methods give very different estimates:
$U_{0}\sim 10^{6}\,K k_{B}$ \cite{aziz1991,szalewicz2012} and
$U_{0}\sim 10^{3}\,K k_{B}$ \cite{rovenchak2000,mt2005},
respectively. From whence, we obtain $\bar{r}_{c}\approx 20 d_{0}$
and $\bar{r}_{c}\approx 2 d_{0}$. The requirement $n\geq n_{c}$
yields $\bar{r} \leq \bar{r}_{c}\approx (2\div 20) d_{0}$. Such
densities correspond to a fluid, a crystal or a dense gas. In this
case, Bogoliubov's criterion \cite{bog1947} is not satisfied. We
note that the magnitude and the sign of the scattering length $a$
can be varied with the help of the Feshbach resonance
\cite{pit2006}.

For a periodic system, $\textbf{k}$ is quantized: $\textbf{k}=2\pi
\left (\frac{l_{x}}{L_{x}}, \frac{l_{y}}{L_{y}}, \frac{l_{z}}{L_{z}}
\right )$. Let $\textbf{k}=2\pi \left (\frac{j_{x}}{L_{x}},
\frac{j_{y}}{L_{y}}, \frac{j_{z}}{L_{z}} \right )$ for the solutions
(\ref{f-n7}) and (\ref{f-n11}). Then there exists the smallest
vector $\textbf{s}=\left (\frac{L_{x}}{|j_{x}|},
\frac{L_{y}}{|j_{y}|}, \frac{L_{z}}{|j_{z}|} \right )$, for which
$\hat{\psi}(\textbf{r}+\textbf{s},t)=\hat{\psi}(\textbf{r},t)$ for
any $\textbf{r}$ (the last equality holds for any of the components
of the vector $\textbf{s}$ as well). We have obtained a
\textit{one-dimensional crystal-like} solution. Indeed, let us put
the axis $x$ along $\textbf{k}$. Then formula (\ref{f-n4}) takes the
form $\hat{\psi}(\textbf{r},t)
=V^{-1/2}(\hat{a}_{0}+\hat{a}_{k}e^{ikx}+\hat{a}_{-k}e^{-ikx})$, and
for the two-particle density matrix
$F_{2}(\textbf{r}_{1},\textbf{r}_{2}|\textbf{r}_{1},\textbf{r}_{2})
=  const\langle
\Psi|\hat{\psi}^{+}(\textbf{r}_{1},t)\hat{\psi}^{+}(\textbf{r}_{2},t)
\hat{\psi}(\textbf{r}_{1},t)\hat{\psi}(\textbf{r}_{2},t)|\Psi\rangle
$ we get
$F_{2}(\textbf{r}_{1},\textbf{r}_{2}|\textbf{r}_{1},\textbf{r}_{2})
=  const
[N_{-k}(N_{-k}-1)+N_{k}(N_{k}-1)+2N_{-k}N_{k}(1+\cos{[2k(x_{1}-x_{2})]})+
2N_{0}(N_{-k}+N_{k})(1+\cos{[k(x_{1}-x_{2})]})+N_{0}(N_{0}-1)]$.
This function has two periods and depends only on the coordinates
$x_{1}$, $x_{2}$. By setting $N_{0}=0$ or $N_{-k}=0$ in this
formula, we obtain
$F_{2}(\textbf{r}_{1},\textbf{r}_{2}|\textbf{r}_{1},\textbf{r}_{2})$
for solutions (\ref{f-n7}) or (\ref{f-n11}), respectively. These are
1D solutions with one period.
%Мы получили одномерное
%\textit{кристаллоподобное} решение. Действительно, направим ось $x$
%вдоль $\textbf{k}$. Тогда формула (\ref{f-n4}) принимает вид
%$\hat{\psi}(\textbf{r},t)
%=V^{-1/2}(\hat{a}_{0}+\hat{a}_{k}e^{ikx}+\hat{a}_{-k}e^{-ikx})$, и
%для парной корреляционной функции
%$F_{2}(\textbf{r}_{1},\textbf{r}_{2}|\textbf{r}_{1},\textbf{r}_{2})
%=  const\langle
%\Psi|\hat{\psi}^{+}(\textbf{r}_{1},t)\hat{\psi}^{+}(\textbf{r}_{2},t)
%\hat{\psi}(\textbf{r}_{1},t)\hat{\psi}(\textbf{r}_{2},t)|\Psi\rangle
%$ находим $F_{2}(\textbf{r}_{1},\textbf{r}_{2}|\textbf{r}_{1},\textbf{r}_{2})
%=  const [N_{-k}(N_{-k}-1)+N_{k}(N_{k}-1)+2N_{-k}N_{k}(1+\cos{[2k(x_{1}-x_{2})]})+
%2N_{0}(N_{-k}+N_{k})(1+\cos{[k(x_{1}-x_{2})]})+N_{0}(N_{0}-1)]$. Это
%функция с двумя периодами, зависящая только от координат   $x_{1}$,
%$x_{2}$. Если положить в этой формуле $N_{0}=0$ или $N_{-k}=0$,
%получим $F_{2}(\textbf{r}_{1},\textbf{r}_{2}|\textbf{r}_{1},\textbf{r}_{2})$
%для решений (\ref{f-n7}) или (\ref{f-n11}) соответственно. Это 1D
%решения с одним периодом.
% We have obtained \textit{a
%crystal-like solution } with a cell of sizes $s_{x},s_{y},s_{z}$. In
%such crystals, one or two components of the vector $\textbf{s}$ can
%be much larger than the mean interatomic distance.
The 1D and 2D systems can be considered similarly. If the ground
state of a natural crystal does contain a condensate, its structure
is seen from the formula
$\hat{\psi}(\textbf{r}+\textbf{s},t)=\hat{\psi}(\textbf{r},t)$ and
the corresponding expansion of the operator
$\hat{\psi}(\textbf{r},t)$ in basis functions. We may expect that,
for periodic boundary conditions (BCs),  the principal harmonic of
the condensate is characterized by the wave vector $\textbf{k}=2\pi
\left (\frac{1}{s_{x}}, \frac{1}{s_{y}}, \frac{1}{s_{z}} \right )$.

Interestingly, our crystal-like solution corresponds to a constant
density. Moreover, it is easy to show that any pure stationary state
of a \textit{periodic} system of spinless particles is characterized
by a constant density. Indeed, let $\hat{\psi}(\textbf{r},t)
=V^{-1/2}\sum_{\textbf{k}}\hat{a}_{\textbf{k}}e^{i\textbf{k}\textbf{r}}$.
Then
\begin{eqnarray}
n(\textbf{r})=\langle
\hat{\psi}^{+}(\textbf{r},t)\hat{\psi}(\textbf{r},t)
\rangle=V^{-1}\sum\limits_{\textbf{k}}\langle
\hat{a}^{+}_{\textbf{k}}\hat{a}_{\textbf{k}}
\rangle=V^{-1}\sum\limits_{\textbf{k}}N_{\textbf{k}}=N/V.
     \label{ro-ex} \end{eqnarray}
In this case, the crystalline properties should be manifested in the
two-particle density matrix
$F_{2}(\textbf{r}_{1},\textbf{r}_{2}|\textbf{r}_{1},\textbf{r}_{2})$
and in the structural factor $S(k)$.

\begin{figure*}
\includegraphics[width=.6\textwidth]{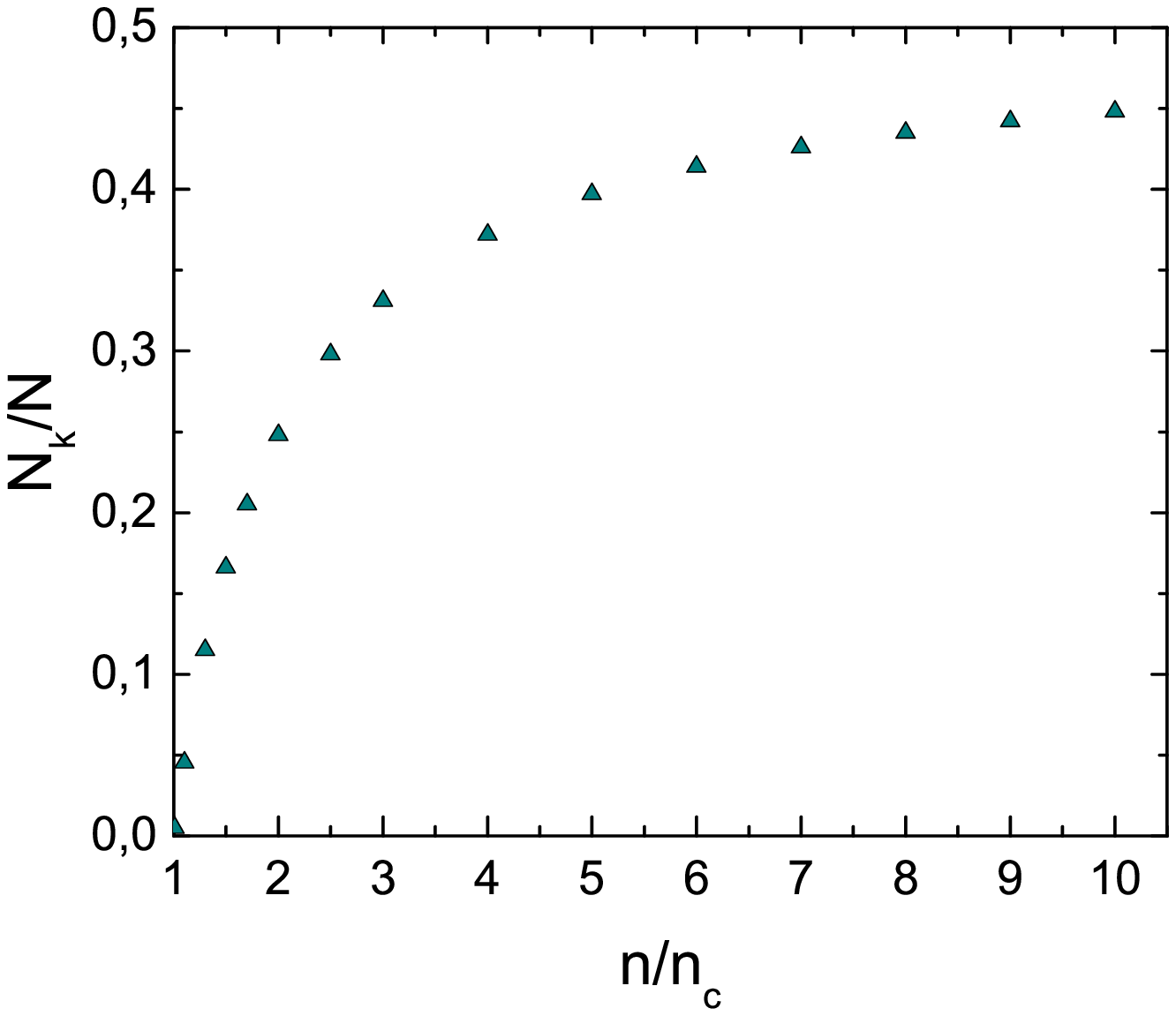}
\caption{[Color online] Values of $N_{\textbf{k}}/N$ corresponding
to the smallest value of $E^{(1+1)}/E^{(1)}$ at the given density
$n$. The solutions for the $1D$ and $3D$ cases are presented (they
coincide for each $n$, in the limits of errors). The smallest value
of $E^{(1+1)}/E^{(1)}$ is determined numerically by means of the
comparison of the values of $E^{(1+1)}(N_{\textbf{k}}, k)/E^{(1)}$,
obtained from Eq. (\ref{f-n16}), for different $k$.}
          \label{fig:3}                  \end{figure*}

Can we observe the fragmented condensate experimentally? We showed
above that the fragmented condensate in the 1D case corresponds to a
highly excited state and, therefore, can hardly be produced. In the
3D case, the periodic BCs are not possible. For zero BCs, the basis
functions are sines. Therefore, the degeneracy
$e^{i\textbf{k}\textbf{r}} \leftrightarrow
e^{-i\textbf{k}\textbf{r}}$ is removed, and the condensate $(0+2)$
should be replaced by a single (nonfragmented) condensate $(0+1)$.
However, the two-condensate state $(1+1)$  should conserve its
structure under zero BCs as well. If such state is sufficiently
close to the ground one, it should be \textit{observable}.
Unfortunately, we do not know whether this state, with regard for
the necessary corrections considered in the following sections, is
close to the ground one. But our above estimates do not forbid the
latter. In Figs. 2 and 3 we present the smallest value (\ref{f-n16})
of the function $E^{(1+1)}(N_{\textbf{k}}, k)$  and the
corresponding $N_{\textbf{k}}/N$ (\ref{nmin2}) for the given density
(at $n= n_{c}$ the smallest $E^{(1+1)}(N_{\textbf{k}}, k)$
corresponds to $g=g_{c}$; the value of $g$ increases insignificantly
with $n$; here, $g=kd_{0}$). At large $n$ the quantity
$N_{\textbf{k}}$ approaches the asymptotic value $N_{\textbf{k}}=
N/2$. Apparently, our conclusions are qualitatively valid also for
the atoms in a harmonic trap.

We note that the crystal-like solutions were previously obtained
numerically for the ground state of a 1D system of dipolar bosons
\cite{lozovik2005,citro2007,citro2008,reimann2010,zollner2011,lode2018}.
The crystallization occurs at the densities exceeding some critical
value. In this case, the field of a trap was considered
\cite{reimann2010,lode2018} or was not
\cite{lozovik2005,citro2007,citro2008,zollner2011}.
%In works \cite{reimann2010,zollner2011}, the dipole-dipole potential is
%three-dimensional. Therefore, the density can be nonuniform even for
%the ring geometry.
Note the interesting comparison of the solutions
for  point and dipolar interatomic interactions  which was executed
in \cite{lode2018b} for strong coupling.  The main difference of the
solutions in
\cite{lozovik2005,citro2007,citro2008,reimann2010,zollner2011,lode2018,lode2018b}
from the above-obtained ones consists in that our solution contains
a condensate. The ground state in works
\cite{citro2008,reimann2010,lode2018}, where the occupation numbers
were calculated, \textit{does not} contain a condensate. This
difference is probably related to the circumstance that our 1D
solutions correspond to highly excited states, whereas the authors
of works
\cite{lozovik2005,citro2007,citro2008,reimann2010,zollner2011,lode2018}
ascribed the solutions to the ground state. According to the theorem
of nodes, if the wave function of the ground state has no nodes and
corresponds to a crystal
\cite{lozovik2005,citro2007,citro2008,reimann2010,zollner2011,lode2018},
then a highly excited state with a lot of nodes and a similar
crystal structure must exist. Our 1D solutions should correspond to
it.

The Pollock-Nozieres' results \cite{pollock1967,nozieres1995} are
important. However, the above analysis shows that, for a
high-density uniform periodic system, Pollock-Nozi\'{e}res' argument
does not work: the fragmentation of a condensate in such system is
possible.

\section{Periodic Bose system: collective description}
\label{sec:3} In Section 2 we described a system of $N$ interacting
bosons with the quasi-single-particle (Hartree--Fock) wave functions
of the form
 \begin{equation}
\Psi_{\{n_{\textbf{k}_{f}}\}}=const
(\hat{a}^{+}_{\textbf{k}_{1}})^{n_{\textbf{k}_{1}}}(\hat{a}^{+}_{\textbf{k}_{2}})^{n_{\textbf{k}_{2}}}\cdots
|vac\rangle, \label{psi000}     \end{equation}
 where
 $n_{\textbf{k}_{1}}+n_{\textbf{k}_{2}}+\ldots =N$ and  $\{n_{\textbf{k}_{f}}\}\equiv (n_{\textbf{k}_{1}},
n_{\textbf{k}_{2}}, \ldots)$. The key point consists in that such
wave functions are not eigenfunctions of the exact Hamiltonian
(\ref{f-0}). Therefore, the energies obtained in Sect. 2 are not
eigenenergies. Indeed, Hamiltonian (\ref{f-0}) can be written in the
form
\begin{eqnarray}
\hat{H}&=&\sum\limits_{\textbf{q}}K(q)\hat{a}^{+}_{\textbf{q}}\hat{a}_{\textbf{q}}+
\sum\limits_{\textbf{k}\textbf{q}_{1}\textbf{q}_{2}}\frac{\nu(k)}{2V}\hat{a}^{+}_{\textbf{k}+\textbf{q}_{1}}
\hat{a}^{+}_{-\textbf{k}+\textbf{q}_{2}}\hat{a}_{\textbf{q}_{1}}\hat{a}_{\textbf{q}_{2}}\nonumber \\
&=&\sum\limits_{\textbf{q}}K(q)\hat{N}_{\textbf{q}}+\sum\limits_{\textbf{q}}\frac{\nu(0)}{2V}\hat{N}_{\textbf{q}}
(\hat{N}_{\textbf{q}}-1)+
 \sum\limits_{\textbf{k}\textbf{q}}^{\textbf{k}\neq 0}\frac{\nu(k)}{2V}\hat{N}_{\textbf{k}+\textbf{q}}
\hat{N}_{\textbf{q}}\nonumber \\ &+&
\sum\limits_{\textbf{q}_{1}\textbf{q}_{2}}^{\textbf{q}_{1}\neq\textbf{q}_{2}}\frac{\nu(0)}{2V}\hat{N}_{\textbf{q}_{1}}
\hat{N}_{\textbf{q}_{2}}+\sum\limits_{\textbf{k}\textbf{q}_{1}\textbf{q}_{2}}^{\textbf{k}\neq
0, \
\textbf{q}_{1}\neq\textbf{q}_{2}}\frac{\nu(k)}{2V}\hat{a}^{+}_{\textbf{k}+\textbf{q}_{1}}\hat{a}_{\textbf{q}_{1}}
\hat{a}^{+}_{\textbf{q}_{2}}\hat{a}_{\textbf{k}+\textbf{q}_{2}}.
     \label{f-n17} \end{eqnarray}
First four terms on the right-hand side of (\ref{f-n17}) do not
change function (\ref{psi000}). But the last term transfers this
function into a superposition of the infinite number of various
terms of the form (\ref{psi000}).  It means that the
quasi-single-particle approach allows one to approximately study the
possibility of the fragmentation of the condensate, but it does not
allow one to find the ground state of the system. We need a more
subtle method allowing one to determine the eigenfunctions and
eigenenergies of the Hamiltonian.

In this section, we propose such method and consider one example of
a solution with fragmented condensate.

The above analysis shows that a part of atoms must be outside the
condensates. Therefore, one needs to consider the harmonics
$\hat{a}_{\textbf{k}}$ with all possible $\textbf{k}$ in the
operator $\hat{\psi}(\textbf{r},t)$ and the Hamiltonian $\hat{H}$
(\ref{f-0}), (\ref{f-n17}). However, in the crude approximation it
is allowable to consider that all atoms of the system are in one or
several condensates. In this case, the wave functions should be
eigenfunctions of the truncated Hamiltonian written in the
corresponding approximation for $\hat{\psi}(\textbf{r},t)$. To
obtain such functions, we use the Landau idea \cite{landau1941}
according to which the weakly excited states of a system of many
interacting particles can be described in the language of
noninteracting quasiparticles. This means that the exact Hamiltonian
(\ref{f-0}), (\ref{f-n17}) must be reduced to the diagonal form
\begin{eqnarray}
 \hat{H}=E_{0}+
 \sum\limits_{\textbf{k}}E(\textbf{k})\hat{\xi}^{+}_{\textbf{k}}\hat{\xi}_{\textbf{k}}.
      \label{h-diag} \end{eqnarray}
In this case, the eigenfunctions of the Hamiltonian take the form
\begin{eqnarray}
 \Psi_{\{n_{\textbf{k}}\}}=C(\hat{\xi}^{+}_{\textbf{k}_{1}})^{n_{\textbf{k}_{1}}}
\ldots
(\hat{\xi}^{+}_{\textbf{k}_{p}})^{n_{\textbf{k}_{p}}}\Psi_{0}.
      \label{e-psi} \end{eqnarray}
Here, $\Psi_{0}$ is the wave function for the state without
quasiparticles, $\hat{\xi}^{+}_{\textbf{k}}$ and
$\hat{\xi}_{\textbf{k}}$ are the operators of creation and
annihilation of a quasiparticle,  and $n_{\textbf{k}_{j}}$ is the
number of quasiparticles with quantum number $\textbf{k}_{j}$. It is
clear that
$\hat{H}\Psi_{\{n_{\textbf{k}}\}}=E_{\{n_{\textbf{k}}\}}\Psi_{\{n_{\textbf{k}}\}}$,
where
$E_{\{n_{\textbf{k}}\}}=E_{0}+\sum_{\textbf{k}}n_{\textbf{k}}E(\textbf{k})$.
Such method allows one to find the operator structure of
eigenfunctions and the eigenenergies $E_{\{n_{\textbf{k}}\}}$ for
lowest levels accurately.

The analysis below is carried on in such a way that the wave
functions are eigenfunctions of the Hamiltonian. For a Bose gas
under periodic BCs, we now compare two states: (i) the state, in
which each of $N$ atoms has the zero momentum, and (ii) the state,
in which $N_{0}, N_{\textbf{k}},$ and $N_{-\textbf{k}}$ atoms have
the momenta $0, \textbf{k},$ and $-\textbf{k},$ respectively (in
this case, $N_{\textbf{k}}, N_{-\textbf{k}}\neq 0$ and
$N_{0}+N_{\textbf{k}}+ N_{-\textbf{k}}=N$).  For the state (i) we
have the wave  function
$\Psi=C_{1}(\hat{a}^{+}_{0})^{N}|vac\rangle$, which is an
eigenfunction of the Hamiltonian $\hat{H}^{(1)}$ (\ref{f-1}) with
the eigenenergy $E^{(1)}$ (\ref{f-1}). For the state (ii) let
$\hat{\psi}(\textbf{r},t)
=V^{-1/2}(\hat{a}_{0}+\hat{a}_{\textbf{k}}e^{i\textbf{k}\textbf{r}}+\hat{a}_{-\textbf{k}}e^{-i\textbf{k}\textbf{r}})$
and $\hat{N}_{\textbf{k}}, \hat{N}_{-\textbf{k}}\ll \hat{N}$ (the
latter condition is necessary for the diagonalization of the
Hamiltonian). The numbers $N_{\textbf{k}}$ and $N_{-\textbf{k}}$ can
be macroscopic or microscopic. The solution for the Hamiltonian is
given by formula (\ref{f-3}), where we neglect the term $\sim
\hat{N}_{\textbf{k}}\hat{N}_{-\textbf{k}}$. We also make
replacements $\hat{a}_{\pm\textbf{k}}\rightarrow
e^{-i\epsilon_{0}t/\hbar}\hat{b}_{\pm\textbf{k}}$,
$\hat{a}_{0}\rightarrow e^{-i\epsilon_{0}t/\hbar}b_{0}$. In the
approximation $\hat{N}_{\textbf{k}}, \hat{N}_{-\textbf{k}}\ll
\hat{N}$ we have $\hat{N}_{0}\approx N_{0}, \hat{N}\approx N$. Then
relation (\ref{f-3}) leads to the Bogoliubov formulae
\cite{bog1947}:
\begin{eqnarray}
&\hat{H}^{(1+2)}&\approx\frac{N_{0}n_{0}\nu(0)}{2}+
[K(\textbf{k})+n_{0}\nu(\textbf{k})+
n_{0}\nu(0)]\hat{b}^{+}_{\textbf{k}}\hat{b}_{\textbf{k}}+
[K(\textbf{-k})+n_{0}\nu(-\textbf{k})+n_{0}\nu(0)]\hat{b}^{+}_{-\textbf{k}}\hat{b}_{-\textbf{k}}\nonumber
\\ &+&
\frac{b_{0}^{2}}{2V}[\nu(\textbf{k})\hat{b}^{+}_{\textbf{k}}\hat{b}^{+}_{-\textbf{k}}+
\nu(-\textbf{k})\hat{b}^{+}_{-\textbf{k}}\hat{b}^{+}_{\textbf{k}}]+
\frac{(b^{*}_{0})^{2}}{2V}[\nu(\textbf{k})\hat{b}_{\textbf{k}}\hat{b}_{-\textbf{k}}+
\nu(-\textbf{k})\hat{b}_{-\textbf{k}}\hat{b}_{\textbf{k}}]
\label{f-4}
\\ &=&\frac{N_{0}n_{0}\nu(0)}{2}+(\hat{N}-\hat{N}_{0})n_{0}\nu(0)+
E(k)-K(k)-n_{0}\nu(k)+E(k)\hat{\xi}^{+}_{\textbf{k}}\hat{\xi}_{\textbf{k}}+
E(k)\hat{\xi}^{+}_{-\textbf{k}}\hat{\xi}_{-\textbf{k}},
   \nonumber   \end{eqnarray}
where $E(k)=\sqrt{K^{2}(k)+2n_{0}\nu(k)K(k)}$ \cite{bog1947}. Using
the eigenfunctions (\ref{e-psi}), we now find the ground-state
energy  in the quasiparticle representation
\cite{bog1947,bog1949,mtmb2017}  as the statistical average $\langle
\hat{H}^{(1+2)} \rangle$ over the state without quasiparticles:
\begin{eqnarray}
E_{0}^{(1+2)}=E_{0}^{(1)}-A(k),
     \label{f-5} \end{eqnarray}
\begin{eqnarray}
A(k)=\frac{(N-N_{0})(n-n_{0})\nu(0)}{2}+
K(k)+n_{0}\nu(k)-E(k)\approx K(k)+n_{0}\nu(k)-E(k),
     \label{f-6} \end{eqnarray}
where $E_{0}^{(1)}=\frac{Nn\nu(0)}{2}$ is the energy of the system,
in which all atoms are in the condensate $\psi(\textbf{r},t)
=V^{-1/2}a_{0}$. In the calculation of $\langle \hat{H}^{(1+2)}
\rangle,$ we considered $N$ to be fixed and used the Gibbs canonical
distribution. For $K(k)+n_{0}\nu(k)> |n_{0}\nu(k)|$ we have $A(k)>0$
and $E_{0}^{(1+2)}<E_{0}^{(1)}$. Therefore, the \textit{fragmented
condensate is possible}.

These solutions imply that, at $K(k)+n_{0}\nu(k)> |n_{0}\nu(k)|,$
the condensate should be fragmented, and the numbers
$N_{\textbf{k}}$, $N_{-\textbf{k}}$ can be macroscopic.
%In addition, at $K(k)+n\nu(k)< 0$ (condition of fragmentation from Sect. 2),
%$A(k)$ becomes complex-valued. Therefore, the fragmentation must
%disappear. However, these conclusions are not quite physical, since
However, more accurate analysis requires the consideration of all
$\textbf{k}$-harmonics. In this case, the Hamiltonian
$\hat{H}^{(1+2)}$ (\ref{f-4}) transits in the known Bogoliubov
Hamiltonian \cite{bog1947}.  Bogoliubov formulae for the equilibrium
occupation numbers $N_{\textbf{k}}=\langle
\hat{a}^{+}_{\textbf{k}}\hat{a}_{\textbf{k}} \rangle$ and
$N_{-\textbf{k}}=\langle
\hat{a}^{+}_{-\textbf{k}}\hat{a}_{-\textbf{k}} \rangle$ imply that
\textit{the numbers $N_{\textbf{k}}$ and $N_{-\textbf{k}}$ can be
macroscopic only for a 1D system}, see also \cite{mtjltp2016}.

Moreover, the Bogoliubov energy $E_{0}$ of the ground state
satisfies the inequality  $E_{0}<E_{0}^{(1)}$. Therefore, we
conclude that it is energetically favorable for a weakly interacting
Bose system with fixed $N$ that a part of atoms has a nonzero
momentum.

Next, in Sect. 2 we noted that $\nu(k)$ can be negative. Despite
this, the Bogoliubov solution satisfies the inequality
$K(k)+n_{0}\nu(k)> |n_{0}\nu(k)|$ for all $\textbf{k}$, because the
Bogoliubov model works at small $|n_{0}\nu(0)|$ \cite{bog1947}, and
since $|\nu(k)|\leq |\nu(0)|$ for any realistic potential. Since
$n_{0} \approx n$, and since $|\nu(k)|$ is not small for $k\lsim
1/d_{0}$ (where $d_{0}$ is the size of an atom), the smallness of
the quantity $|n_{0}\nu(k)|$ means the smallness of $n$. Thus, the
analysis in Sect. 3 is applicable only to systems with low density.
Such analysis cannot verify the validity of the solutions with
fragmentation from Sect. 2, since these solutions correspond to a
high density ($n>n_{c}$), which breaks the Bogoliubov criterion
\cite{bog1947}.

To verify the validity of the crystal-like solutions with fragmented
condensate, which are obtained in Sect. 2, it is necessary to
diagonalize the Hamiltonian for a condensate of corresponding
structure, by considering all $\textbf{k}$-harmonics and preserving
the terms $\sim \hat{N}_{\textbf{k}}\hat{N}_{-\textbf{k}}$. Since
the functions (\ref{f-n6}) and (\ref{f-n10}) are eigenfunctions of
the corresponding truncated Hamiltonians, it is quite probable that
the exact condition of fragmentation is close to the condition
$K(k)+n\nu(k)< 0$ obtained in Sect. 2.

Furthermore, it follows from the formula
\begin{equation}
\hat{\psi}(\textbf{r},t)
=V^{-1/2}e^{-i\epsilon_{0}t/\hbar}(\hat{b}_{0}+\hat{b}_{\textbf{k}}e^{i\textbf{k}\textbf{r}}+\hat{b}_{-\textbf{k}}e^{-i\textbf{k}\textbf{r}})
\label{f-unif} \end{equation}  that the system is uniform:
\begin{equation}
 n(\textbf{r})=\langle
\hat{\psi}^{+}(\textbf{r},t)\hat{\psi}(\textbf{r},t) \rangle=\langle
\hat{b}^{+}_{0}\hat{b}_{0}+
\hat{b}^{+}_{\textbf{k}}\hat{b}_{\textbf{k}}+\hat{b}^{+}_{-\textbf{k}}\hat{b}_{-\textbf{k}}+
\hat{b}^{+}_{\textbf{k}}\hat{b}_{-\textbf{k}}e^{-2i\textbf{k}\textit{r}}+
\hat{b}^{+}_{-\textbf{k}}\hat{b}_{\textbf{k}}e^{2i\textbf{k}\textit{r}}\rangle
/V=N/V.
     \nonumber \end{equation}
Here, we used the Bogoliubov transformations, which yield $\langle
\hat{b}^{+}_{\textbf{k}}\hat{b}_{-\textbf{k}}\rangle=\langle
\hat{b}^{+}_{-\textbf{k}}\hat{b}_{\textbf{k}}\rangle=0$.

The Bogoliubov method \cite{bog1947,bog1949} allows one to describe
the weakly excited states of an equilibrium Bose system.
% (see also the discussion of the Bogoliubov method in \cite{mtmb2017}).
Note that the method works for sufficiently large $N$: $N\gsim
N_{cr}$. For a 1D system, the Bogoliubov solutions
\cite{bog1947,mtmb2017} agree with the exact ones
\cite{ll1963,lieb1963,gaudin1971,mt2015,mtsp2018} at $N\gsim 100$
under periodic BCs and at $N\gsim 1000$ under zero BCs. Therefore,
$N_{cr}\simeq 100$ for periodic BCs, and $N_{cr}\simeq 1000$ for the
zero ones.

We note that, for real gases in a trap, it is necessary to consider
quasiparticles and the variability of the number of particles. In
this case, one needs to average over the grand canonical ensemble
\cite{huang}.

We mention the work by Nozi\'{e}res and Saint James
\cite{nozieres1982}, where a pair condensation and a fragmentation
of the condensate were studied within the variation method
considering the finite size of particles and the anomalous averages.
This method differs significantly from our one. In this case, a
solution with a fragmented condensate was not found in
\cite{nozieres1982}.

We also mention the interesting work by Streltsov
\cite{streltsov2013}, in which it was shown that the ground state of
a 1D Bose gas is fragmented, if the repulsive interatomic
interaction is strong and the interaction radius is comparable with
the system size. Our analysis in Sect. 3 is valid only at weak
coupling. But the solutions in Sect. 2 are applicable in the case of
strong coupling and hint that the fragmentation found in
\cite{streltsov2013} is related to the nonpointness of atoms.

Thus, in this section we have studied the solution $(1+2)$
(\ref{f-3})--(\ref{f-3e}) from Sect. 2 within a more accurate
approach. We have required additionally that $N_{\textbf{k}},
N_{-\textbf{k}}\neq 0$ and $N_{\textbf{k}}, N_{-\textbf{k}}\ll N$,
which prohibits solutions $(0+2)$ and $(1+1)$ from Sect. 2. With the
account for all $\textbf{k}$-harmonics, our analysis yields the
Bogoliubov Hamiltonian. Therefore, it is necessary to use
Bogoliubov's criterion for the density \cite{bog1947}, which gives
$n\ll n_{c}$. Under such condition, the inequality $A(k)>0$ holds,
and the fragmentation into three condensates ($0, \textbf{k},
-\textbf{k}$) is energy-gained. However, according to Sect. 2, a
\textit{one-condensate} solution is energy-gained at $n\ll n_{c}$.
In the analysis in Sect. 3, the fragmentation is energy-gained at
$n\ll n_{c}$ due to the ``anomalous'' averages  $\langle
\hat{b}^{+}_{\textbf{k}}\hat{b}^{+}_{-\textbf{k}} \rangle$, $\langle
\hat{b}_{\textbf{k}}\hat{b}_{-\textbf{k}} \rangle$.  In Sect. 2,
instead of the averages $\langle
\hat{b}^{+}_{\textbf{k}}\hat{b}^{+}_{-\textbf{k}} \rangle$, $\langle
\hat{b}_{\textbf{k}}\hat{b}_{-\textbf{k}} \rangle$ we considered the
normal quantum-mechanical average $\langle \hat{H}_{scat} \rangle$,
which is zero in the quasi-single-particle representation. Because
of this, the possibility of a fragmentation for small $n$ was lost
in Sect. 2. If we consider all $\textbf{k}$-harmonics in Sect. 2,
the anomalous averages will not appear nevertheless, since they
arise only within the collective approach. Therefore, the collective
approach is basically more accurate than the quasi-one-particle one.

According to the analysis  in Sect. 3, the fragmentation of the
condensate is possible in a 1D Bose gas at $T=0$ and a weak
coupling. We have found no fragmentation in 2D and 3D Bose gases
(here, the conclusion by Pollock and Nozi\'{e}res is proper).
Interestingly, the condition of fragmentation $K(k)+n\nu(k)< 0$ (see
Sect. 2) obtained in the quasi-single-particle approach is opposite
to the condition $K(k)+n_{0}\nu(k)> |n_{0}\nu(k)|$ following from
the collective approach (Sect. 3). The nonpointness of atoms favors
the fragmentation in the first case and counteracts in the second
one. We note that the condition $K(k)+n_{0}\nu(k)> |n_{0}\nu(k)|$
was obtained for the ground state and the weak coupling, whereas the
condition $K(k)+n\nu(k)< 0$ is true in the case of strong coupling
and  non-ground state.

\section{One-dimensional Bose gas under zero boundary conditions}
\label{sec:4} In Sections 2 and 3 we have found the solutions
containing only three $\textbf{k}$-harmonics. Below, we will
determine the structure of the condensate in the collective approach
involving all $\textbf{k}$-harmonics. We use zero BCs:
$\hat{\psi}(x,t)=0$ at $x=0, L$. A similar problem was solved
numerically in the case of strong coupling at $T=0$, $N\lsim 100$
\cite{streltsov2013}. We will consider analytically a system with
weak coupling, $T\geq 0$, and $N\gsim 1000$. Previously, with the
help of the Bogoliubov method we constructed the description of
weakly excited states of a Bose gas under zero BCs and found the
density matrix $F_{1}(x,x^{\prime})$ \cite{mtmb2017}. We emphasize
that the Bogoliubov method describes \textit{well} a finite $1D$
system at a weak coupling and $T\rightarrow 0$. This follows from
the facts that the criterion of applicability of the method is
satisfied \cite{mtmb2017}, the solutions for $E_{0}$ and $E(k)$
coincide with the solutions in the exactly solvable approach based
on the Bethe ansatz
\cite{ll1963,lieb1963,gaudin1971,mt2015,mtsp2018}, and the solution
for $F_{1}(x,x^{\prime})|_{T=0}$ is close to the solution for a
periodic system, obtained by different methods (see references in
\cite{mtmb2017}). The solution for the density matrix of a 1D Bose
gas under zero BCs reads \cite{mtmb2017}:
\begin{eqnarray}
F_{1}(x,x^{\prime})&=& \tilde{F}_{1}(x,x^{\prime})
 +\sum\limits_{l=1,2,\ldots,\infty}
\chi_{2l}\varphi_{2l}^{*}(x^{\prime})\varphi_{2l}(x),
     \label{f-7} \end{eqnarray}
\begin{eqnarray}
\tilde{F}_{1}(x,x^{\prime})=
f_{0}^{*}(x^{\prime})f_{0}(x)+\frac{2}{L}\sum\limits_{j=1,2,\ldots}
\chi_{2j-1}\sin{(k_{2j-1}x^{\prime})}\sin{(k_{2j-1}x)},
     \label{f-8} \end{eqnarray}
 \begin{eqnarray}
\chi_{j} =\frac{1}{\sqrt{y_{j}^{4}+4y_{j}^{2}}}\left
(\frac{2}{\sqrt{y_{j}^{4}+4y^{2}_{j}}+y^{2}_{j}+2}+\frac{y^{2}_{j}+2}{e^{\frac{\sqrt{y_{j}^{4}+4y_{j}^{2}}}{\tilde{T}}}-1}
\right ),
      \label{4-12} \end{eqnarray}
\begin{equation}
f_{0}(x)=
\frac{4\sqrt{n_{0}}}{\pi}\sum\limits_{j=1,2,\ldots,\infty}\frac{\sin{(k_{2j-1}x)}}{2j-1}
\frac{4}{y^{2}_{2j-1}+4},
     \label{5-15} \end{equation}
where $L$ is the size of the system, $k_{j}=\frac{\pi j}{L}$,
$\varphi_{2l}(x)=\sqrt{\frac{2}{L}}\cdot\sin{(k_{2l}x)}$,
$y_{j}=\frac{j}{\sqrt{\Gamma}}$, $\Gamma=\frac{\gamma
NN_{0}}{\pi^{2}}$, $\tilde{T}=\frac{k_{B}T}{cn_{0}}$, and
$n_{0}=\frac{N_{0}}{L}$. The solution is written for the point
interatomic interaction [$U(|x_{j}-x_{l}|)=2c\delta(x_{j}-x_{l})$,
$\gamma=\frac{2mc}{\hbar^{2}n}$, $n=\frac{N}{L}$] and is valid for
$0<\gamma\ll 1$, $\Gamma \gg 1$, $N_{0}\approx N\gsim 10^{3}$.  The
point approximation is justified for the description of states with
$k_{j}\ll \pi/r_{0}$, since the transition to a potential with
nonzero radius  $r_{0}$ changes such solutions slightly.

It is seen from (\ref{f-7}) and (\ref{f-8}) that the expansion of
the function $\tilde{F}_{1}(x,x^{\prime})$ is nondiagonal, but the
sum $\sum_{l} \chi_{2l}\varphi_{2l}^{*}(x^{\prime})\varphi_{2l}(x)$
has a diagonal form. In this case, $\tilde{F}_{1}(x,x^{\prime})$ is
orthogonal (in each of the arguments $x$ and $x^{\prime}$) to any
term of the sum $\sum_{l}
\chi_{2l}\varphi_{2l}^{*}(x^{\prime})\varphi_{2l}(x)$, and the
functions $\varphi_{2l}(x)$ are orthonormalized. Therefore, it is
clear that $\sum_{l}
\chi_{2l}\varphi_{2l}^{*}(x^{\prime})\varphi_{2l}(x)$ is the sum
$\sum_{l} \lambda_{2l}\varphi_{2l}^{*}(x^{\prime})\varphi_{2l}(x)$
from the diagonal expansion (\ref{n2}).  To represent the function
$F_{1}(x,x^{\prime})$ in the form (\ref{n2}), we need to find a
diagonal expansion
\begin{equation}
\tilde{F}_{1}(x,x^{\prime})=
 \sum\limits_{j=1,2,\ldots,\infty}
\lambda_{2j-1}\varphi_{2j-1}^{*}(x^{\prime})\varphi_{2j-1}(x).
     \label{f-9} \end{equation}
It is convenient to pass from (\ref{f-9}) to the equivalent system
of equations
\begin{equation}
\int\limits_{0}^{L}
dx^{\prime}\varphi_{2j-1}(x^{\prime})\tilde{F}_{1}(x,x^{\prime})=
 \lambda_{2j-1}\varphi_{2j-1}(x), \quad j=1,2,\ldots,\infty.
     \label{f-10} \end{equation}
We seek the functions $\varphi_{2j-1}(x)$ in the form
\begin{equation}
\varphi_{2j-1}(x)=\sum\limits_{l=1,2,\ldots,\infty}A_{2l-1}^{(2j-1)}\sqrt{\frac{2}{L}}\sin{(k_{2l-1}x)},
     \label{f-11} \end{equation}
which ensures the orthogonality of $\varphi_{2j-1}(x)$ to the
functions $\varphi_{2l}(x)$. Let us substitute (\ref{f-11}) in
(\ref{f-10}) and take formulae (\ref{f-8}), (\ref{5-15}) into
account. We obtain the system of equations
\begin{eqnarray}
&&\sum\limits_{l=1,2,\ldots,\infty}A_{2l-1}^{(2j-1)}(\chi_{2l-1}-\lambda_{2j-1})\sin{(k_{2l-1}x)}\nonumber
\\ &+&
\frac{8N_{0}}{\pi^{2}}\sum\limits_{p,l=1,2,\ldots,\infty}\frac{A_{2p-1}^{(2j-1)}}{2p-1}\frac{\sin{(k_{2l-1}x)}}{2l-1}
\frac{4}{4+y_{2p-1}^{2}}\frac{4}{4+y_{2l-1}^{2}}=0,
     \label{f-12} \end{eqnarray}
where $j=1,2,\ldots,\infty$. By equating the coefficients of the
functions $\sin{(k_{2l-1}x)}$ to zero, we get
\begin{equation}
A_{2l-1}^{(2j-1)}=-\frac{8N_{0}}{\pi^{2}}\frac{1}{2l-1}\frac{4}{4+y_{2l-1}^{2}}\frac{S_{2j-1}}{\chi_{2l-1}-\lambda_{2j-1}},
\quad j,l=1,2,\ldots,\infty,
     \label{f-13} \end{equation}
\begin{equation}
S_{2j-1}=\sum\limits_{l=1,2,\ldots,\infty}\frac{A_{2l-1}^{(2j-1)}}{2l-1}\frac{4}{4+y_{2l-1}^{2}}.
     \label{f-14} \end{equation}
Substituting $A_{2l-1}^{(2j-1)}$ in (\ref{f-14}), we obtain the
secular equation for the numbers $\lambda_{2j-1}$:
\begin{equation}
1+\sum\limits_{l=1,2,\ldots,\infty}\frac{f_{2l-1}}{\chi_{2l-1}-\lambda_{2j-1}}=0,
\quad
f_{2l-1}=\frac{8N_{0}}{\pi^{2}}\frac{1}{(2l-1)^{2}}\frac{4^{2}}{(4+y_{2l-1}^{2})^{2}}.
     \label{f-15} \end{equation}
It is easy to show analytically that $\lambda_{1}\approx N_{0}$ (for
$\gamma \ll 1$) and $\lambda_{2j-1}\in ]\chi_{2j-1},\chi_{2j-3}[$
for $j\geq 2$.

We  note that, for the interacting system, the genuine condensate is
determined by the diagonal expansion (\ref{n2}), where the number
$\lambda_{j}/N$ is the probability of the location of an atom in the
single-particle state $\phi_{j}(\textbf{r})$. The average $\langle
\hat{\psi}(x,t)\rangle$ is also often called a condensate. Usually,
$\langle \hat{\psi}(x,t)\rangle$ coincides with the condensate
determined with the help of (\ref{n2}). But such a coincidence is
not always the case (see below). Therefore, we will call the
quantity $\langle \hat{\psi}(x,t)\rangle$ the effective condensate.

The density matrix $F_{1}(x,x+x^{\prime})$ (\ref{f-7}) at $T=0$
decreases, as $|x^{\prime}|$ increases, by a power law
$|x^{\prime}|^{-|s|}$ with $s=\sqrt{\gamma}/2\pi$ \cite{mtmb2017}.
In this case, it is accepted to talk about a quasicondensate instead
of a condensate (fragmented or not). The Bogoliubov method works at
$|s|\ll 1$. Therefore, for a finite system,
$F_{1}(x,x+x^{\prime})\approx const$ for all points $x^{\prime}$ not
too close to boundaries (see details in \cite{mtmb2017}). In this
case, the quasicondensate can be considered as a \textit{true}
condensate.  For the infinite system,
$F_{1}(x,x+x^{\prime})|_{x^{\prime}\rightarrow \infty} =const \cdot
|x^{\prime}|^{-|s|}\rightarrow 0$ even for very small nonzero $|s|$.
We arrive at Hohenberg's conclusion that the condensate is absent
\cite{hohenberg1967}. Thus, the true condensate can exist in a $1D$
Bose system, if this system is finite.

\subsection{The case of $T=0$}
We now present the solutions $\lambda_{j}$ for $\Gamma=10^{7}$,
$N=10^{5}$, see Table 1. We have checked this solution. It satisfies
the normalization $\lambda_{1}+\lambda_{2}+\ldots +\lambda_{50001}=
0.999N$, and the functions $\varphi_{2j-1}(x)$ (\ref{f-11}),
(\ref{f-13}) are orthogonal to each other. Since
$F_{1}(x,x^{\prime})=F^{*}_{1}(x^{\prime},x)$, the eigenvalues
$\lambda_{l}$ in (\ref{n2}) are real, and the collection
$\{\lambda_{l}\}$ is unique \cite{korn}. In addition, if all
$\lambda_{l}$ are different, the natural basis $\{\phi_{l}(x)\}$ is
unique \cite{korn}. In our case, all $\lambda_{l}$ are different.
Therefore, the above solution is unique. Note that the functions
$\varphi_{2j+1}(x)$ are roughly close to
$-\sqrt{2/L}\cdot\cos{k_{2j}x}$. At different $\Gamma, N$, we have
$\lambda_{2j+1}< \lambda_{2j}$ provided that $j\geq 1$. Thus, we
have found the diagonal expansion (\ref{n2}).

\begin{table}
      % \vskip 3mm
      \begin{center}
\caption{\large \ Natural occupations  $\lambda_{j}$ for different
$\Gamma$,  $N$, and $\tilde{T}=\frac{k_{B}T}{cn_{0}}$.   We
determined the values of $q_{0}$, $q_{T}$, $\tilde{N}_{0}$, $N_{0}$,
and $\gamma$ from the formula $\Gamma=\frac{\gamma N_{0}N}{\pi^{2}}$
and Eqs. (78)--(83) in work \cite{mtmb2017}. Here, $\tilde{N}_{0}$
is the number of atoms in the effective condensate $\langle
\hat{\psi}(x,t)\rangle$, and $N_{0}$, $q_{0}$, $q_{T}$ are auxiliary
numbers \cite{mtmb2017}. The numbers $\lambda_{2j+1}$ were obtained
by solving Eq. (\ref{f-15}) numerically. For the ``even'' harmonics
we have $\lambda_{2l}=\chi_{2l}$ (\ref{4-12}). }  \vskip 10mm
 \begin{tabular}{|c|c|c|c|c|c|c|c|}  \hline
   $\tilde{T}$     & 0           & 0.0005 & 0.001      & 0         &  0       & 0.01     & 0.02   \\ \hline
   $\Gamma$        & $10^{7}$    & $10^{7}$ & $10^{7}$ & $10^{6}$  & $10^{6}$ & $10^{6}$ & $10^{6}$ \\ \hline
   $N$          & $10^{5}$  & $10^{5}$ & $10^{5}$ & $3.5\cdot 10^{4}$& $10^{5}$ & $10^{5}$ & $10^{5}$ \\ \hline
   $q_{0}$       & 0.995492      & 0.995492 & 0.995492  & 0.99479 & 0.99479   & 0.99479  & 0.99479 \\ \hline
   $q_{T}$       &               & 0.34422 & 0.550123  &         &           & 0.791791 & 0.876522 \\ \hline
   $\tilde{N}_{0}/N$ & 0.87315   & 0.859037 & 0.82804  & 0.90183   & 0.965641    & 0.900714 & 0.821891 \\ \hline
   $N_{0}/N$       & 0.87328     & 0.859166 & 0.828165 & 0.90226   & 0.966102    & 0.901145 & 0.822284 \\ \hline
   $\gamma$        & 0.011302      & 0.011487 & 0.011917 & 0.00893  & 0.001022  & 0.001095 & 0.0012 \\ \hline
   $\lambda_{1}/N$ & 0.886652     & 0.882725 & 0.871153 & 0.91402  & 0.969903  & 0.942379 & 0.905786 \\ \hline
   $\lambda_{2}/N$ & 0.0079      & 0.009269 & 0.014119 & 0.00713  & 0.002495  & 0.012661 & 0.025078 \\ \hline
   $\lambda_{3}/N$ & 0.0066      & 0.008741 & 0.013613 & 0.00596  & 0.00209   & 0.01231  & 0.023645 \\ \hline
   $\lambda_{4}/N$ & 0.00395     & 0.003998 & 0.004632 & 0.00356  & 0.001245  & 0.003285 & 0.006328  \\ \hline
   $\lambda_{5}/N$ & 0.00354     & 0.003752 & 0.0045   & 0.00318  & 0.001115  & 0.003256 & 0.006228 \\ \hline
   $\lambda_{6}/N$ & 0.00263     & 0.002633 & 0.002751 & 0.00237  & 0.000828  & 0.001547 & 0.002856 \\ \hline
   $\lambda_{7}/N$ & 0.00242     & 0.002487 & 0.002666 & 0.00218  & 0.000764  & 0.001537 & 0.002833  \\ \hline
   $\lambda_{8}/N$ & 0.00197     & 0.001972 & 0.001967 & 0.00177  & 0.00062   & 0.000936 & 0.00164 \\ \hline
   $\lambda_{9}/N$ & 0.00185     & 0.001877 & 0.001935 & 0.00166  & 0.000581  & 0.00093  & 0.001632 \\ \hline
   $\lambda_{10}/N$ & 0.00158    & 0.001576 & 0.001582 & 0.00141  & 0.000495  & 0.000651 & 0.001077 \\ \hline
   $\lambda_{11}/N$ & 0.00149    & 0.00151  & 0.001535 & 0.00134  & 0.000469  & 0.000647 & 0.001073 \\ \hline
   $\lambda_{50}/N$ & 0.000311   & 0.000311 & 0.000311 &          & 0.000095  & 0.000095 & 0.000096 \\ \hline
   $\lambda_{51}/N$ & 0.000295   & 0.000307 & 0.000308 &          & 0.00009   & 0.000094 & 0.000095 \\ \hline
   $\lambda_{100}/N$ & 0.000153  & 0.000153 & 0.000153 &          & 0.000045  & 0.000045 & 0.000045  \\ \hline
   $\lambda_{101}/N$ & 0.000149  & 0.000148 & 0.000152 &          & 0.000044  & 0.000045 & 0.000045 \\ \hline
         \end{tabular} \end{center} \end{table}

The above solution has two significant properties.  (I) The
quasicondensate can be fragmented. Indeed, for a finite system we
may consider the state $\varphi_{j}(x)$ to be macroscopically
occupied at $\lambda_{j}\gsim N/\Theta$. Here, the choice of the
value of $\Theta$ is somewhat arbitrary. Whether $\lambda_{j}=0.03N$
is macroscopic? Probably not if $N\lsim 100$. Probably yes if
$N\gsim 10^{4}$. In our opinion, it is reasonable to set
$\Theta=(\ln{N})^{2}$. According to such criterion, states $2$ and
$3$ from the above solution (for $\tilde{T}=0$, $\Gamma=10^{7}$,
$N=10^{5}$) are occupied macroscopically. (II) The structure of a
fragmented quasicondensate depends on the boundaries. Indeed, it is
easy to obtain from the Bogoliubov formulae \cite{bog1947} that, for
a periodic system,
\begin{eqnarray}
F_{1}(x,x^{\prime})
&=&\sum\limits\limits_{j=1,2,\ldots}\chi_{-2j}\phi^{p*}_{-2j}(x^{\prime})\phi^{p}_{-2j}(x)+
N_{0}\phi^{p*}_{0}(x^{\prime})\phi^{p}_{0}(x)\nonumber
\\&+&\sum\limits\limits_{j=1,2,\ldots}\chi_{2j}\phi^{p*}_{2j}(x^{\prime})\phi^{p}_{2j}(x),
     \label{f-16} \end{eqnarray}
where $\phi^{p}_{2j}(x)=e^{ik_{2j}x}/\sqrt{L}$, and
$\chi_{-2j}=\chi_{2j}$ is set by formula (\ref{4-12}). We remark
that for a periodic system
$F_{1}(x,x^{\prime})=F_{1}(x-x^{\prime})$, and the Fourier transform
of the function $F_{1}(x-x^{\prime})$ coincides with (\ref{f-16}).
The solution $F_{1}(x,x^{\prime})$ obtained above under zero BCs can
be written in a similar way:
\begin{eqnarray}
F_{1}(x,x^{\prime})
&=&\sum\limits\limits_{j=1,2,\ldots}\lambda_{2j+1}\phi^{*}_{2j+1}(x^{\prime})\phi_{2j+1}(x)+
\lambda_{1}\phi^{*}_{1}(x^{\prime})\phi_{1}(x)\nonumber
\\&+&\sum\limits\limits_{j=1,2,\ldots}\lambda_{2j}\phi^{*}_{2j}(x^{\prime})\phi_{2j}(x).
     \label{f-17} \end{eqnarray}
Here, $\lambda_{1}\approx N$ and
$\lambda_{2j}=\chi_{2j}\neq\lambda_{2j+1}$. Thus, under  periodic
BCs we have $\lambda_{-2j}=\lambda_{2j}$. However, under  zero BCs
the analogous symmetry is absent: $\lambda_{2j+1}\neq\lambda_{2j}$.
The difference between $\lambda_{2j+1}$ and $\lambda_{2j}$ is
essential for small $j$ and decreases, as $j$ increases. The
property $\lambda_{-2j}=\lambda_{2j}$ is related to the cyclic
symmetry of the system. The boundaries break this symmetry;
therefore, the equality $\lambda_{2j+1}=\lambda_{2j}$ is also
violated. Thus, a change in the numbers $\lambda_{j}$ at the
transition from periodic BCs to the zero ones is related to a change
in the topology of the system.

For the system under  zero BCs we now clarify the conditions, under
which the quasicondensate is fragmented. At small $l$ we have
$\lambda_{2l}=\chi_{2l}\simeq\frac{1}{2y_{2l}}=\frac{\sqrt{\Gamma}}{4l}\approx
\frac{N\sqrt{\gamma}}{4\pi l}$ (here, we have used that
$N_{0}\approx N$ at the weak coupling \cite{mtmb2017}). In this
case, $\lambda_{2l+1}=\lambda_{2l}-|\delta_{2l}|$, where
$\delta_{2l}$ is small. The criterion $\lambda_{2l}\gsim
\frac{N}{(\ln{N})^{2}}$ requires $\sqrt{\gamma}\gsim \frac{4\pi
l}{(\ln{N})^{2}}$. These formulae imply that the states $2,3,4,
\ldots, 2l+1$ are macroscopically occupied, if
\begin{equation}
\sqrt{\gamma} \gsim \frac{4\pi l}{(\ln{N})^{2}}.
     \label{f-18} \end{equation}
On the other hand, the criterion of applicability of the Bogoliubov
method, $N-\tilde{N}_{0}\lsim 0.1N,$ and the formulae $N_{0}\approx
\tilde{N}_{0}$,  $1-\frac{\tilde{N}_{0}}{N}\approx
\frac{\sqrt{\Gamma}}{4N}\ln{\Gamma}$ \cite{mtmb2017} yield the
inequality
\begin{equation}
\sqrt{\gamma} \lsim \frac{0.4\pi}{\ln{(\gamma N^{2}/\pi^{2})}}.
     \label{f-19} \end{equation}
Inequalities (\ref{f-18}) and (\ref{f-19}) are compatible only for
definite values of $\gamma$ and $N$.  In particular, for  $N\lsim
10^{3}$ inequalities (\ref{f-18}) and (\ref{f-19}) are not
compatible. For $N= 10^{4}$ they are compatible, if $\gamma\approx
0.015$,  $l=1$ (in this case, the states $1, 2, 3$ are
macroscopically occupied). For $N= 10^{5}$ we find $\gamma\approx
0.01$, $l=1$. If $N= 10^{10}$, then the inequalities are compatible
for $6\cdot 10^{-4}\lsim \gamma< 2\cdot 10^{-3}$, $l=1$ and for
$\gamma\simeq 2\cdot 10^{-3}$, $l=2$ (in the last case, the states
$1, 2, 3, 4, 5$ are macroscopically filled). We do not consider the
values $N> 10^{10},$ since they are not experimentally realizable.

The diagonal representation (\ref{f-16}) for a periodic 1D Bose
system at $T=0$ was found previously by a different method
\cite{mtjltp2016}. Instead of $\chi_{2l}$ (\ref{4-12}), close
occupation numbers were obtained:
\begin{equation}
\lambda_{2l}=\frac{\sqrt{\gamma}N_{0}}{4|l|\pi}, \quad l=\pm 1, \pm
2, \ldots
      \label{f-21} \end{equation}
This formula holds for $l\ll \sqrt{\Gamma}$. At the replacement
$N_{0}\rightarrow \sqrt{NN_{0}}$ formula (\ref{f-21}) passes to
$\lambda_{2l}=\frac{\sqrt{\Gamma}}{4|l|}$, which coincides with
$\chi_{2l}$ (\ref{4-12}) at  $\tilde{T}=0$, $l\ll \sqrt{\Gamma}$.
The difference between $N_{0}$ and $\sqrt{NN_{0}}$ is insignificant,
since the methods in \cite{mtjltp2016,mtmb2017} require
$N_{0}\approx N$. Note that the density matrix was found in
\cite{mtjltp2016} directly from the ground-state wave function
without any assumptions about the condensate. At $\gamma \lsim 0.01$
the solution in \cite{mtjltp2016} is close to the exact one.

It is clear that, as $\gamma$ increases, the atoms from the lowest
single-particle states transit in higher ones. Therefore, we may
expect that the number of lowest macroscopically populated states
increases with $\gamma$. At $\gamma\gg 1$ the atoms are apparently
distributed over the very large number of states, and there are no
macroscopically occupied states. However, we cannot verify these
assumptions, since the methods in \cite{bog1947,mtjltp2016,mtmb2017}
are valid only at small $\gamma$.

As we noted above, the condensate exists only in a \textit{finite}
1D system. Bogoliubov's method is also applicable only to a finite
(1D) system (condition (\ref{f-19})). The quasicondensate
(condensate) is fragmented, if condition (\ref{f-18}) with $l\geq 1$
is satisfied. Inequality (\ref{f-18}) follows from the criterion
$\lambda_{2l}=\chi_{2l}\gsim \frac{N}{(\ln{N})^{2}}$ and formula
(\ref{4-12}) for the quantity $\chi_{2l}\equiv N_{k_{2l}}$
\cite{mtmb2017}. Since the occupation numbers $N_{k_{j}}$ at $T=0$
should correspond to the smallest energy of the system, inequality
(\ref{f-18}) is, in fact, the condition for the fragmentation of a
condensate to be energy-gained.

\subsection{The case of $T>0$}
The thermal equilibrium in a system is possible, if the number of
quasiparticles is large. This requires \cite{mtmb2017} that
$E(k_{1})\ll k_{B}T$, which yields $\tilde{T}\gg
y_{1}=\Gamma^{-1/2}\approx \frac{\pi}{\sqrt{\gamma}N}$ (here, $E(k)$
is the dispersion law of quasiparticles). On the other hand, the
criterion of applicability of the Bogoliubov method
$0<\frac{\sqrt{\gamma}}{2\pi}\ln{\frac{N\sqrt{\gamma}}{\pi}}+0.08\gamma
N\tilde{T}\ll 1$  \cite{mtmb2017} requires $\tilde{T}\ll
\frac{12}{\gamma N}$.  In this case, for $\sqrt{\Gamma}\gg 1$ and
small $j$, relation (\ref{4-12}) yields
\begin{equation}
\chi_{j}\approx \frac{1}{2y_{j}}\left
(1+\frac{2}{e^{\frac{2y_{j}}{\tilde{T}}}-1} \right )\approx
\frac{1}{2y_{j}}\left (1+\frac{\tilde{T}}{y_{j}} \right ).
     \label{f-22} \end{equation}
If $j\lsim 10$, then $\tilde{T}\gg y_{j}$. Therefore, the main
contribution to $\chi_{j}$ is given by the temperature term
$\tilde{T}/y_{j}$. Thus, at $ y_{1} \ll\tilde{T}\ll \frac{12}{\gamma
N}$ the temperature affects  the density matrix significantly.

In Table 1 we present the solutions with the above-considered
parameters $\Gamma=10^{7}$, $N=10^{5}$ for $\tilde{T}=0.0005;
0.001$. At both temperatures, the states $1, 2, 3$ are filled
macroscopically.

Let us consider the case $\Gamma=10^{6}$, $N=10^{5}$ for
$\tilde{T}=0; 0.01; 0.02$. As is seen from Table 1, at $\tilde{T}=0$
only the state $1$ is macroscopically occupied. At $\tilde{T}=0.01$,
the states $1, 2, 3$ are macroscopically  populated. At last, for
$\tilde{T}=0.02$ the states $1, 2, 3, 4, 5$ are macroscopically
occupied.

We see that, as $\tilde{T}$ increases, the atoms transit from the
state $1$ to the states $2$, $3$ and to higher ones. It cannot be
excluded that, at sufficiently high temperatures, the state $1$ is
occupied microscopically, but the states $2$ and $3$ are occupied
macroscopically.

Interestingly, for a finite system the order parameter $\langle
\hat{\psi}(x,t)\rangle$ does not generally coincide with the genuine
condensate defined with the help of criterion (\ref{n2}). Under
periodic BCs, the function $F_{1}(x,x^{\prime})$ is set by formula
(\ref{f-16}), and the number $\tilde{N}_{0}$  of atoms in the
effective condensate $\langle \hat{\psi}(x,t)\rangle$ is equal to
$N_{0}$. If the genuine condensate is not fragmented, it coincides
with $\langle \hat{\psi}(x,t)\rangle$. But if the genuine condensate
is fragmented, there is no coincidence, since the states
$\phi^{p}_{\pm 2}(x), \phi^{p}_{\pm 4}(x), \ldots$ do not enter the
average $\langle \hat{\psi}(x,t)\rangle =const \cdot
e^{-i\epsilon_{0}t/\hbar}$. Under  zero BCs, the effective
condensate $\langle \hat{\psi}(x,t)\rangle$ does not coincide with
the genuine one, since $\tilde{N}_{0} \neq \lambda_{1}$ even if the
genuine condensate is not fragmented.  For example, for $\Gamma
=10^{6}$, $N=3.5\cdot 10^{4}$, $T=0$ we get $\lambda_{1}\approx
0.914N$, $\lambda_{2}\approx 0.00713N$, $\lambda_{3}\approx
0.00596N$ (see Table 1). According to the criterion
$\lambda_{j}\gsim N/(\ln{N})^{2}$, only the state $1$ is
macroscopically occupied. In this case,
$\tilde{N}_{0}\neq\lambda_{1}$. This noncoincidence is related to
the anomalous averages and the difference of the natural occupations
$\lambda_{j}$ under the zero  and periodic BCs. For periodic BCs,
$\tilde{N}_{0}=\lambda_{1}=N_{0}$ (the states $-2, -4, \ldots$ under
periodic BCs correspond to the states $3, 5, \ldots$ under zero BCs;
at the transition from the periodic to zero BCs, a part of atoms
passes from the states $-2, -4, \ldots$ to the state $1$ for zero
BCs). However, even if the effective condensate does not coincide
with the genuine one, the former is close to the latter, at least
for the weak coupling. For the applicability of the Bogoliubov
method to a 1D Bose system, namely the effective condensate $\langle
\hat{\psi}(x,t)\rangle$ is significant: The number of atoms
$\tilde{N}_{0}$ in this condensate should be close to $N$
\cite{mtmb2017}.

\section{Conclusion}
\label{sec:5} We have shown in two ways that the fragmentation of
the condensate in a uniform Bose system is possible. Within the
quasi-single-particle approach, we have found approximate
one-dimensional crystal-like solutions with a fragmented condensate.
Such solutions are possible for 1D, 2D, and 3D high-density system.
However,  they apparently correspond to highly excited states of the
system.  With the help of the more accurate collective approach, we
obtained that the ground state of a uniform 1D Bose system with
repulsive interatomic potential contains a fragmented
quasicondensate at low $T$  and at definite values of the parameters
of the system. In this case, the number of quasicondensates forming
a fragmented quasicondensate can be equal to 3 or 5. The occupation
numbers of a fragmented quasicondensate depend on the boundary
conditions, though the energy of the ground state $E_{0}$ and the
dispersion law $E(k)$ are independent of BCs
\cite{ll1963,mtmb2017,gaudin1971,mtsp2018}. In recent years, the
experiments with a uniform gas in a trap became possible
\cite{lopes2017}. Therefore, we hope for that the above obtained
solutions will be verified experimentally.

\textit{Note added in proof.} Recently, we became aware of works
\cite{gross1958,kirz}, in which crystal-like solutions with a
condensate of atoms were also considered.

%\begin{acknowledgements}
The present work was partially supported by the Program of
Fundamental Research of the Department of Physics and Astronomy of
the National Academy of Sciences of Ukraine (project No.
0117U000240).
% выше - это целевая
%\end{acknowledgements}

%Отбросить шестой знак в 3м и 4м столбцах табл.

    \renewcommand\refname{}

%\begin{table}
%\begin{tabular}{lll}
%\hline\noalign{\smallskip}
%\end{tabular}
%\end{table}

% On page 9, we made an additional correction that is absent in the published version.

       \end{document}